\documentstyle[seceq,epsf]{ptptex}

\def\beqa{\begin{eqnarray}}
\def\eeqa{\end{eqnarray}}
\def\beqn{\begin{eqnarray}}
\def\eeqn{\end{eqnarray}}
\def\beq{\begin{equation}}
\def\eeq{\end{equation}}

\def\O{\Omega}
\def\L{\Lambda}
\def\D{\Delta}
\def\d{\delta}
\def\r{\rho}
\def\a{\alpha}
\def\b{\beta}
\def\g{\gamma}
\def\t{\tilde}

\def\p{\partial}

\def\ra{\rightarrow}
\def\pa{\partial}

\title{
Irrotational and Incompressible Ellipsoids \\
in the First Post-Newtonian Approximation of General Relativity
}

\author{
Keisuke Taniguchi,
\footnote{E-mail: taniguci@tap.scphys.kyoto-u.ac.jp,}~
Hideki Asada
\footnote{E-mail: asada@phys.hirosaki-u.ac.jp}~
and ~
Masaru Shibata
\footnote{E-mail: shibata@vega.ess.sci.osaka-u.ac.jp}
}

\inst{
$^\ast$Department of Physics,~Kyoto University,~Kyoto
  606-8502,~Japan \\
$^{**}$Faculty of Science and Technology,~Hirosaki
  University,~Hirosaki 036-8561,~Japan \\
$^{***}$Department of Earth and Space Science,~Graduate School of
  Science \\
  Osaka University,~Toyonaka,~Osaka 560-0043,~Japan
}

\abst{
First post-Newtonian (1PN) hydrostatic equations for an irrotational
fluid which have been recently derived are solved for an incompressible
star. The 1PN configurations are expressed as a deformation of the
Newtonian irrotational Riemann ellipsoid using Lagrangian displacement
vectors introduced by Chandrasekhar. For the 1PN solutions, we also
calculate the luminosity of gravitational waves in the 1PN approximation
using the Blanchet-Damour formalism. It is found that the solutions of
the 1PN equations exhibit singularities at points where the axial ratios
of semi-axes are $1:0.5244:0.6579$ and $1:0.2374:0.2963$, and the
singularities seem to show that at the points, the irrotational Riemann 
ellipsoid is unstable to the deformation induced by the effect of
general relativity. For stable cases ($a_2/a_1 > 0.5244$, where
$a_1$ and $a_2$ are the semi-major and minor axes, respectively) we find
that when increasing the 1PN correction, the angular velocity and total
angular momentum increase, while the total energy and luminosity of
gravitational waves decrease. These 1PN solutions will be useful when
examining the accuracy of numerical code for obtaining relativistic
irrotational stars.

We also investigate the validity of an ellipsoidal approximation, in
which a 1PN solution is obtained assuming an ellipsoidal figure and
neglecting the deformation. It is found that for $a_2/a_1 > 0.7$, the
ellipsoidal approximation gives a fairly accurate result for
the energy, angular momentum, and angular velocity,
although in the approximation we cannot find the singularities.
}

\begin{document}
\maketitle

\section{Introduction}

Preparation of reliable theoretical models on the late inspiraling stage
of binary neutron stars is one of the most important issues for
gravitational wave astronomy. This is because they represent one
promising source for gravitational wave detectors such as
LIGO,\cite{LIGO} VIRGO,\cite{VIRGO} GEO600\cite{GEO} and
TAMA.\cite{TAMA} From their signals, we will obtain a wide variety of
physical information on neutron stars such as their mass and spin, if we
have a theoretical template of them.\cite{KIP} In particular, a signal
from the very late inspiraling stage just prior to merging may contain
physically important information on neutron stars such as their
radius,\cite{KIP} which will be utilized for determining the equation of 
state of neutron stars.\cite{lindblom}

Binary neutron stars evolve due to the radiation reaction of
gravitational waves, so that they never settle down to equilibrium
states. However, the emission time scale will always be longer than the
orbital period outside their innermost stable circular orbit (ISCO), so
that we may consider them to be in quasiequilibrium states in their
inspiraling phase even near the ISCO.

Until now, all the reliable relativistic works devoted to obtaining a
quasiequilibrium state have been undertaken assuming a corotational
velocity field,\cite{shiba,BCSST} since there was no formalism to
compute non-corotational solutions. As pointed out previously,
\cite{Kochanek,BC} however, corotation is not an adequate assumption for
the velocity field of realistic binary neutron stars, because the effect
of viscosity is negligible for the evolution of neutron stars in a
binary system and, as a result, their velocity fields are expected to be
irrotational (or nearly irrotational). Formalism for the computation of
realistic quasiequilibrium states of coalescing binary neutron stars
just prior to merging has recently been developed by several authors in
general relativity.\cite{BGM,Asada,Shibata,Teukolsky,Gourgoulhon} In
this formalism, we have only to solve two hydrostatic equations as for
the fluid equations. One of them is the integrated form of the Euler
equation and the other is the Poisson equation for the velocity
potential. Thus, the formalism seems to be very tractable for computing
equilibrium configurations of relativistic irrotational bodies.

In this paper, we apply the formalism for solving an incompressible,
irrotational single star in the first post-Newtonian (1PN) approximation
as a first step. Incompressible rotating stars were studied extensively
at the 1PN order by Chandrasekhar about 30 years ago. He derived many
equilibrium configurations of the Maclaurin, Jacobi, Dedekind
ellipsoids, \cite{Ch65,chandra65,chandra67,chandra74} developing
original methods (see also Ref. 20)), and he found many interesting
features of rotating stars deformed by relativistic effects.

The purpose of this paper is twofold. One is to develop a formalism to
obtain an equilibrium state of an irrotational, incompressible fluid in
the 1PN approximation. As mentioned above, Chandrasekhar studied 1PN
equilibrium stars extensively, but he did not do so for irrotational
stars. We develop the method in this paper which can be applied even for
irrotational binary systems. The other comes from a demand in performing
numerical computation: Our final aim is to study general relativistic
irrotational binary neutron stars with compressible equations of state.
Thus, numerical computation is necessary. Although some promising
numerical methods are proposed,\cite{UE,BGM} the computation still does
not seem easy. Hence, when we obtain a result by numerical computation,
it is necessary to check its validity by comparing with an analytic
solution. Thus, preparation of exact solutions is required.

This paper is organized as follows. In \S 2, we describe the basic
equations to calculate equilibrium configurations of the 1PN
irrotational star. The deformation of the star from an ellipsoid and the
angular velocity in the 1PN approximation are calculated in \S 3.  In \S
4, we give the boundary conditions to determine the velocity field and
the deformation in the 1PN approximation, and the total energy and
angular momentum are obtained in \S 5. We present formalism for
computation of gravitational radiation from a rotating star with
arbitrary internal motion in \S 6. In \S 7, we present the numerical
results obtained by solving the equations derived in previous sections.
We also compare the results with those in the ellipsoidal approximation.
\cite{LRS} Section 8 is devoted to summary and discussion.

Throughout this paper, $c$ denotes the light velocity and we use the
units in which $G=1$. Latin indices $i,j,k, \cdots$ take values 1 to 3,
and $\d_{ij}$ denotes the Kronecker delta.

\section{Formulation}

Non-axisymmetric equilibrium configurations with non-uniform velocity
fields are obtained by solving the Euler, continuity, and Poisson
equations, consistently. Since we consider an incompressible fluid,
all the calculations are carried out analytically even in the 1PN case.
The procedure is as follows.

\vskip 1mm
\noindent
(1) We calculate a Newtonian equilibrium state, i.e., the
irrotational Riemann ellipsoid,\cite{chandra69} as a non-perturbed 
state. For simplicity, we consider only the case in which the directions 
of the vorticity vector and the angular velocity vector 
lie along $x_3$-axis. 

\vskip 1mm
\noindent
(2) 1PN corrections for the velocity potential and gravitational 
potentials are obtained from the 1PN Poisson equations for them. 

\vskip 1mm
\noindent
(3) We calculate the deformation from the Newtonian ellipsoid 
induced by 1PN gravity using Lagrangian displacement vectors 
introduced by Chandrasekhar.\cite{chandra69}
Then, corrections of the Newtonian quantities due to the 
deformation of the star are estimated. 

\vskip 1mm
\noindent
(4) We substitute all the 1PN corrections obtained in (2) and (3) 
into the 1PN Euler and continuity equations. Then, coefficients of 
the Lagrangian displacement vectors and 1PN velocity potential are 
determined from the boundary conditions on the stellar surface. 

\vskip 1mm
In this section, we calculate the Newtonian and 1PN terms which 
we need in the above procedures. In the following, we assume that the
center of mass of the star is located at the origin of the coordinate
system, and the direction of the coordinate axes are parallel to the
principal axes of the star, whose lengths are defined as $a_1, a_2$ and
$a_3$. 

\subsection{Hydrostatic and Poisson equations for 1PN irrotational stars}

For an irrotational fluid, 
the relativistic Euler equation can be integrated, and 
in the 1PN case, it is written as \cite{Shibata}
\beqa
  {\rm const}&=&{P \over \r} -U +{1 \over 2} \sum_k (\p_k
  \phi_{\rm N})^2 -\sum_k \ell^k \p_k \phi_{\rm N} \nonumber \\
  & &+ {1 \over c^2} \biggl[ -{P \over \r} U +{1 \over 2} U^2 +X -{1 \over
    2} \Bigl( {P \over \r} +3U \Bigr) \sum_k (\p_k \phi_{\rm N})^2 -{1
    \over 8} \Bigl( \sum_k (\p_k \phi_{\rm N})^2 \Bigr)^2 \nonumber \\
  & &\hspace{30pt} +\sum_k ( \p_k \phi_{\rm N} ) 
  ( \p_k \phi_{\rm PN} ) -\sum_k \ell^k \p_k \phi_{\rm PN} -\sum_k 
  \hat{\b}^k \p_k \phi_{\rm N} \biggr], \label{euler}
\eeqa
where we assume that fluid is incompressible, i.e., $\rho=$const. 
In Eq. (\ref{euler}), 
$P$, $\r$, $\ell^k$, $\phi_{\rm N}$, $\phi_{\rm PN}$, $U$, $X$, and
$\hat{\b}^k$, respectively, 
denote the pressure, the density, 
the velocity field of the figure rotation (spatial component 
of the Killing vector), 
the Newtonian and 1PN velocity potentials, and the last 
three terms are the Newtonian and 1PN potentials, which are 
derived by solving Poisson equations as 
\beqa
&&\D U=-4\pi \r, \\
&&\D X=4 \pi \r \Bigl[ 2U +{3P \over \r} +2 \sum_k (\p_k \phi_{\rm N})^2 
  \Bigr], \\
&&\D P_k=-4\pi \r \p_k \phi_{\rm N},  \\
&&\D \chi=4\pi \r \sum_k (\p_k \phi_{\rm N}) x^k.
\eeqa 
Here $\hat{\b}_k$ is expressed as
\beqa
  \hat{\b}_k =-{7 \over 2} P_k +
{1 \over 2} \bigl( \p_k \chi +\sum_l  x^l \p_k  P_l \bigr). 
\eeqa
We note that using the gravitational potentials, the spacetime 
line element to 1PN order can be written as
\beq
ds^2=-\alpha^2 c^2 dt^2 + {2 \over c^2} \sum_i \hat \beta_i dx^i dt + 
\Bigl(1+{2U \over c^2}\Bigr) \sum_i dx^i dx^i,
\eeq
where
\beq
\alpha= 1 -{U \over c^2} +{1 \over c^4} \Bigl( {U^2 \over 2}+X\Bigr)
+O(c^{-6}). 
\eeq

A characteristic feature of the irrotational fluid is that 
the continuity equation reduces to a 
Poisson type equation for a velocity potential $\phi$,
\cite{Shibata,Teukolsky} and 
in the 1PN incompressible case, it is 
\beq
\sum_i \rho \pa_i C_i=0,
\eeq
where 
\beq
C_i=-\ell^i +\pa_i \phi_{\rm N}
+{1 \over c^2}\biggl[-\ell^i\Bigl({1 \over 2}\sum_k (\pa_k \phi_{\rm N})^2
+3U\Bigr)-{P \over \rho}\pa_i \phi_{\rm N}
-\hat \beta_i + \pa_i \phi_{\rm PN}\biggr] . 
\label{1PNvelocity}
\eeq
In the following subsections, we obtain terms appearing 
in Eq. (\ref{euler}) by separately solving equations for each.

\subsection{Newtonian terms}

The solution of the Newtonian potential for the ellipsoidal star is
\beqa
  U= \pi \r \big( A_0 -\sum_k A_k x_k^2 \bigr),
\eeqa
where
$A_{ij \cdots}$ are index symbols introduced by
Chandrasekhar,\cite{chandra69} and $A_0$ is calculated
from\cite{chandra69}
\beqa
  A_0 &=& a_1 a_2 a_3 \int_0^{\infty} {du \over \sqrt{(a_1^2 +u) (a_2^2
      +u) (a_3^2 +u)}} \nonumber \\
  &=& a_1^2 \a_2 \a_3 \int_0^{\infty} {dt \over \sqrt{(1+t) (\a_2^2 +t)
      (\a_3^2 +t)}},
\eeqa
where $\a_2 =a_2/a_1$ and $\a_3 =a_3/a_1$.
Also, the pressure at Newtonian order is written as
\beqa
  P= P_0 \Bigl( 1- \sum_k {x_k^2 \over a_k^2} \Bigr),
\eeqa
where $P_0 =\pi \r^2 a_3^2 A_3$ denotes the pressure at the center of
the star.

The Newtonian velocity field in the inertial frame 
$v_i \equiv \pa_i \phi_{\rm N}$ is
written as
\beqa
  v_1&=&\ell^1 +u_1=-\Bigl({a_1^2 f_R \over a_1^2+a_2^2}+1 \Bigr) \O x_2, 
\nonumber \\
  v_2&=&\ell^2 +u_2=\Bigl( {a_2^2 f_R \over a_1^2+a_2^2}+1 \Bigr) \O x_1, 
\nonumber \\
  v_3&=&0, \label{nvelo9}
\eeqa
where $u_i$ is the velocity field in the corotating frame and $\O$
denotes the angular velocity of the figure. $\ell^i$ and $u_i$ are given 
by
\beqn
  \ell^i &=&(-\O x_2, ~\O x_1, ~0), \\
  u_i &=& \Bigl( {a_1 \over a_2} \L x_2, ~-{a_2 \over a_1} \L x_1,~0 
\Bigr)\nonumber \\ 
      &=& \Bigl(-{a_1^2 f_R \over a_1^2+a_2^2} \O x_2, 
                  ~{a_2^2 f_R \over a_1^2+a_2^2} \O x_1, ~0\Bigr), 
\eeqn
where $\L$ is the angular velocity of the internal motion. $f_R$ is
defined as $\zeta/\O$, where $\zeta \equiv ({\rm rot} {\bf u})_3$ 
denotes the vorticity in the corotating frame. 
For the irrotational ellipsoid, $f_R$ becomes $-2$. 
Thus, $\phi_{\rm N}=F_a \O x_1 x_2$ and 
\beq
  v_i=(F_a \O x_2, F_a \O x_1, 0), \label{nvelo}
\eeq
where
\beqa
  F_a \equiv {a_1^2 -a_2^2 \over a_1^2 +a_2^2}.
\eeqa

\subsection{1PN terms}

\subsubsection{$X$}

For the convenience of the calculation, we separate $X$ into two parts as
\beqa
  X=X_0+X_v,
\eeqa
where $X_0$ and $X_v$ are derived from Poisson-like equations as
\beqa
  \D X_0&=&4 \pi \r \Bigl( 2U +{3P \over \r} \Bigr), \\
  \D X_v&=&8 \pi \r \sum_k (\p_k \phi_{\rm N})^2.
\eeqa
Then, the equation for $X_0$ becomes 
\beq
  \D X_0 
  =4\pi \r \biggl[ \Bigl( 2\pi \r A_0 +{3P_0 \over \r} \Bigr) -\sum_k 
  \Bigl( 2\pi \r A_k +{3P_0 \over \r a_k^2} \Bigr) x_k^2 \biggr],
\eeq
and the solution is
\beqa
  X_0 =-\a_0 U +\sum_k \eta_k D_{kk},
\eeqa
where
\beqa
  \a_0&=&2\pi \r A_0 +{3P_0 \over \r}, \\
  \eta_k&=&2\pi \r A_k +{3P_0 \over \r a_k^2}, \\
  D_{kk}&=&\pi \r \Bigl[ a_k^4 \bigl( A_{kk} -\sum_m A_{kkm}
  x_m^2 \bigr) x_k^2 \nonumber \\ 
&& \hskip8mm +{1 \over 4} a_k^2 \bigl( B_k -2\sum_m B_{km} x_m^2
  +\sum_m \sum_n B_{kmn} x_m^2 x_n^2 \bigr) \Bigr],
\eeqa
and $B_{ij\cdots}$ are index symbols introduced by
Chandrasekhar.\cite{chandra69}
$D_{ij \cdots}$ is calculated from a Poisson equation
as\cite{chandra69}
\beqa
  \D D_{ij \cdots} =-4 \pi \r x_i x_j \cdots.
\eeqa

The equation for $X_v$ is
\beq
  \D X_v =8 \pi \r F_a^2 \O^2 (x_1^2 +x_2^2),
\eeq
and hence, the solution is 
\beqa
  X_v=-2 F_a^2 \O^2 (D_{11} +D_{22}).
\eeqa

\subsubsection{$\hat{\b}_k$}

Substituting $v_i$ of Newtonian order  
into the equations for $P_i$ and $\chi$, 
we immediately find the solutions as
\beqa
&&  P_1 =F_a \O D_2,\\
&&  P_2 =F_a \O D_1, \\
&&  P_3 =0, \\
&&  \chi=-2 F_a \O D_{12}, 
\eeqa
where \cite{chandra69}
\beqa
&&  D_i =\pi \r a_i^2 \bigl( A_i -\sum_k A_{ik} x_k^2 \bigr) x_i,\\
&&  D_{12}=\pi \r a_1^2 a_2^2 \bigl( A_{12} -\sum_k A_{12k} x_k^2 \bigr)
  x_1 x_2.
\eeqa
Using the above solutions of $P_k$ and $\chi$, 
we obtain $\hat{\b}_k$:
\beqa
  \hat{\b}_1
  &=&\pi \r F_a \O \biggl[ \Bigl({1 \over 2} a_1^2 A_1 -{7 \over 2}
  a_2^2 A_2 -a_1^2 a_2^2 A_{12} \Bigr) x_2  \nonumber \\
  & &\hspace{40pt} +\Bigl( -{3 \over 2} a_1^2
  A_{11} +{5 \over 2} a_2^2 A_{12} +3a_1^2 a_2^2 A_{112} \Bigr) x_1^2
  x_2 \nonumber \\
  & &\hspace{40pt} +\Bigl( -{1 \over 2} a_1^2 A_{12} +{7 \over 2} a_2^2
  A_{22} +a_1^2 a_2^2 A_{122} \Bigr) x_2^3 \nonumber \\
  & &\hspace{40pt} +\Bigl( -{1 \over 2} a_1^2
  A_{13} +{7 \over 2} a_2^2 A_{23} +a_1^2 a_2^2 A_{123} \Bigr) x_2 x_3^2
  \biggr], \\
  \hat{\b}_2
  &=&\pi \r F_a \O \biggl[ \Bigl({1 \over 2} a_2^2 A_2 -{7 \over 2}
  a_1^2 A_1 -a_1^2 a_2^2 A_{12} \Bigr) x_1 \nonumber \\
  & &\hspace{40pt} +\Bigl( -{1 \over 2} a_2^2
  A_{12} +{7 \over 2} a_1^2 A_{11} +a_1^2 a_2^2 A_{112} \Bigr) x_1^3
  \nonumber \\
  & &\hspace{40pt} +\Bigl( -{3 \over 2} a_2^2 A_{22} +{5 \over 2} a_1^2
  A_{12} +3a_1^2 a_2^2 A_{122} \Bigr) x_1 x_2^2 \nonumber \\
  & &\hspace{40pt} +\Bigl( -{1 \over 2}
  a_2^2 A_{23} +{7 \over 2} a_1^2 A_{13} +a_1^2 a_2^2 A_{123} \Bigr) x_1
  x_3^2 \biggr], \\
  \hat{\b}_3
  &=&\pi \r F_a \O \bigl( 2a_1^2 a_2^2 A_{123} -a_2^2 A_{23} -a_1^2
  A_{13} \bigr) x_1 x_2 x_3.
\eeqa
Then, the divergence of $\hat \beta_k$ is 
\beqa
  \sum_k \p_k \hat{\b}_k
  &=&6 \pi \r F_a \O (a_1^2 +a_2^2) A_{12} x_1 x_2, 
\eeqa
where we have used some relations among index symbols.\cite{chandra69}

\subsubsection{$\phi_{\rm PN}$}

An explicit form of 
the Poisson-type equation for $\phi_{\rm PN}$ can be written as 
\beqa
  &\D \phi_{\rm PN}& =\Bigl[ 6\pi \r \O (A_1 -A_2) 
   -2F_a \O \Bigl( {1 \over a_1^2} +{1 \over a_2^2} \Bigr) {P_0 \over \r} 
\Bigr] x_1 x_2 +\sum_i \p_i
  \hat{\b}_i \nonumber \\
  & &=-2\O {(a_1^2 -a_2^2) \over a_1^2 a_2^2} {P_0 \over \r} x_1 x_2.
\eeqa
The solution of this equation can be written up to biquadratic terms in 
$x_i$ as\footnote{Although we can add higher order terms which satisfy 
$\Delta \phi_{\rm PN}=0$, we negelct them for simplicity because 
we consider only biquadratic deformtion of ellipsoids in this paper.}
\beqa
  \phi_{\rm PN} =(p +q x_1^2 +r x_2^2 +s x_3^2) x_1 x_2
  +{\rm const}, \label{phipn}
\eeqa
where $q$, $r$ and $s$ satisfy the condition
\beqa
  3q +3r +s= -\O {(a_1^2 -a_2^2) \over a_1^2 a_2^2} {P_0 \over
    \r}. \label{cond1}
\eeqa

\subsection{Collection}

Substituting the terms derived above into Eq. (\ref{euler}), we 
obtain  
\beqa
  {\cal G} &\equiv&{P \over \r} -U -\d U +{1 \over 2} F_a^2 \Bigl( \O_R^2
  +{1 \over c^2} \d \O^2 \Bigr) (x_1^2 +x_2^2) \nonumber \\
  & &+F_a \Bigl( \O_R^2 +{1
    \over c^2} \d \O^2 \Bigr) (x_2^2 -x_1^2) \nonumber \\
  & &+{1 \over c^2} \Bigl( \g_0 +\sum_l \g_l x_l^2 +\sum_{l \le m}
  \g_{lm} x_l^2 x_m^2 \Bigr) \nonumber \\
  &=&{\rm const}, \label{integrated}
\eeqa
where $\O_R$ denotes the angular velocity of an irrotational Riemann
ellipsoid and the $\g_{ij}$ are expressed as
\beqa
  \g_0 &=& \pi \r \biggl[ -A_0 \Bigl( {4P_0 \over \r} +{3 \over 2} \pi
  \r A_0 \Bigr) +{1 \over 4} \sum_l \Bigl( 2\pi \r a_l^2 A_l +{3P_0
    \over \r} \Bigr) B_l \nonumber \\
  & &\hspace{20pt}-{1 \over 2} F_a^2 \O_R^2 (a_1^2 B_1 +a_2^2 B_2)
  \biggr], \\
  \g_1 &=& \pi \r \biggl[ 4A_1 {P_0 \over \r} +\pi \r A_0 A_1 +A_0 {P_0
    \over \r a_1^2} \nonumber \\
  & &\hspace{20pt}+\Bigl( 2\pi \r 
  a_1^2 A_1 +{3P_0 \over \r} \Bigr) \Bigl( a_1^2 A_{11} -{1 \over 2}
  B_{11} \Bigr) -{1 \over 2} \Bigl( 2\pi \r a_2^2 A_2 +{3P_0
    \over \r} \Bigr) B_{12}\nonumber \\ 
  & &\hspace{20pt}-{1 \over 2} \Bigl( 2\pi \r a_3^2 A_3 +{3P_0 \over \r}
  \Bigr) B_{13} \nonumber \\
  & &\hspace{20pt} +F_a^2 \O_R^2 ( 3a_1^2 A_1
  -a_2^2 A_2 -2a_3^2 A_3 -3a_1^4 A_{11} ) -{2a_2^2 \over a_1^2 +a_2^2}
  {\O_R \over \pi \r} p \biggr], \\
  \g_2 &=& \pi \r \biggl[ 4A_2 {P_0 \over \r} +\pi \r A_0 A_2 +A_0 {P_0
    \over \r a_2^2} +\Bigl( 2\pi \r a_2^2 A_2 +{3P_0 \over \r} \Bigr)
  \Bigl( a_2^2 A_{22} -{1 \over 2} B_{22} \Bigr) \nonumber \\
  & &\hspace{20pt} -{1 \over 2} \Bigl(
  2\pi \r a_1^2 A_1 +{3P_0 \over \r} \Bigr) B_{12} 
  -{1 \over 2} \Bigl( 2\pi \r a_3^2 A_3 +{3P_0
    \over \r} \Bigr) B_{23} \nonumber \\
  & &\hspace{20pt} +F_a^2 \O_R^2 ( 3a_2^2 A_2 -a_1^2 A_1
  -2a_3^2 A_3 -3a_2^4 A_{22}) +{2a_1^2 \over a_1^2 +a_2^2}
  {\O_R \over \pi \r} p \biggr], \\
  \g_3 &=& \pi \r \biggl[ 4A_3 {P_0 \over \r} +\pi \r A_0 A_3 +A_0 {P_0
    \over \r a_3^2} +\Bigl( 2\pi \r a_3^2 A_3 +{3P_0 \over \r} \Bigr)
  \Bigl( a_3^2 A_{33} -{1 \over 2} B_{33} \Bigr) \nonumber \\
  & &\hspace{20pt} -{1 \over 2} \Bigl(
  2\pi \r a_1^2 A_1 +{3P_0 \over \r} \Bigr) B_{13} 
  -{1 \over 2} \Bigl( 2\pi \r a_2^2 
  A_2 +{3P_0 \over \r} \Bigr) B_{23} \nonumber \\
  & &\hspace{20pt} +F_a^2 \O_R^2 (a_1^2 B_{13}  +a_2^2 B_{23}) \biggr], \\
  \g_{11} &=& \pi \r \biggl[ -A_1 {P_0 \over \r a_1^2}
  +{1 \over 2} \pi \r A_1^2 +\Bigl(
  2\pi \r a_1^2 A_1 +{3P_0 \over \r} \Bigr) \Bigl( -a_1^2 A_{111} +{1
    \over 4} B_{111} \Bigr) \nonumber \\
  & &\hspace{20pt}+{1 \over 4} \Bigl( 2\pi \r a_2^2 A_2 +{3P_0 
    \over \r} \Bigr) B_{112} 
  +{1 \over 4} \Bigl( 2\pi \r a_3^2 A_3 +{3P_0 \over \r}
  \Bigr) B_{113} \nonumber \\
  & &\hspace{20pt} +{F_a^2 \O_R^2 \over 2a_1^2} (3a_1^2 A_1 +a_3^2 A_3
  -8a_1^4 A_{11} +5a_1^6 A_{111} -a_1^4 a_2^2 A_{112}) \nonumber \\
  & &\hspace{20pt}-{F_a^4 \O_R^4 \over 8\pi \r} -{2a_2^2 \over a_1^2
    +a_2^2} {\O_R \over \pi \r} q \biggr], \\
  \g_{12} &=& \pi \r \biggl[ -{P_0 \over \r}
  \Bigl( {A_1 \over a_2^2} +{A_2 \over a_1^2} \Bigr) +\pi \r 
  A_1 A_2 +\Bigl( 2\pi \r a_1^2 A_1 +{3P_0 \over \r} \Bigr) \Bigl(
  -a_1^2 A_{112} +{1 \over 2} B_{112} \Bigr) \nonumber \\
  & &\hspace{20pt}+\Bigl( 2\pi \r a_2^2 A_2
  +{3P_0 \over \r} \Bigr) \Bigl( -a_2^2 A_{122} +{1 \over 2} B_{122}
  \Bigr) +{1 \over 2} \Bigl( 2\pi \r a_3^2 A_3 +{3P_0 \over \r} \Bigr)
  B_{123} \nonumber \\
  & &\hspace{20pt}+{F_a^2 \O_R^2 \over 2} \Bigl( 3A_1 +3A_2 + 
  {a_3^2 (a_1^2 +a_2^2) \over a_1^2 a_2^2} A_3 \nonumber \\ 
  & &\hspace{20pt} -4(a_1^2 + a_2^2) A_{12}
  +3a_1^2 (a_1^2 -a_2^2) A_{112} -3a_2^2 (a_1^2 -a_2^2) A_{122} \Bigr)
  \nonumber \\
  & &\hspace{20pt}-{F_a^4 \O_R^4 \over 4\pi \r} +{6a_1^2 \over a_1^2
    +a_2^2} {\O_R \over \pi \r} q
  -{6a_2^2 \over a_1^2 +a_2^2} {\O_R \over \pi \r} r \biggr], \\
  \g_{13} &=& \pi \r \biggl[ -{P_0 \over \r}
  \Bigl( {A_1 \over a_3^2} +{A_3 \over a_1^2} \Bigr) +\pi \r A_1 A_3
  +\Bigl( 2\pi \r a_1^2 A_1 +{3P_0 \over \r} \Bigr) \Bigl( -a_1^2
  A_{113} +{1 \over 2} B_{113} \Bigr) \nonumber \\
  & &\hspace{20pt} +{1 \over 2} \Bigl( 2\pi \r a_2^2
  A_2 +{3P_0 \over \r} \Bigr) B_{123}
  +\Bigl( 2\pi \r a_3^2 A_3 +{3P_0 \over \r} \Bigr) 
  \Bigl( -a_3^2 A_{133} +{1 \over 2} B_{133} \Bigr)\nonumber \\
  & &\hspace{20pt}+{F_a^2 \O_R^2 \over 2} (4A_3 -a_2^2 A_{23} -9a_1^2
  A_{13} +6a_1^4 A_{113}) 
 -{2a_2^2 \over a_1^2 +a_2^2} {\O_R \over \pi \r} s
  \biggr], \\
  \g_{22} &=& \pi \r \biggl[ -A_2 {P_0 \over \r a_2^2}
  +{1 \over 2} \pi \r A_2^2 +\Bigl(
  2\pi \r a_2^2 A_2 +{3P_0 \over \r} \Bigr) \Bigl( -a_2^2 A_{222} +{1
    \over 4} B_{222} \Bigr) \nonumber \\
  & &\hspace{20pt} +{1 \over 4} \Bigl( 2\pi \r a_1^2 A_1 +{3P_0 
    \over \r} \Bigr) B_{122} 
  +{1 \over 4} \Bigl( 2\pi \r a_3^2 A_3
  +{3P_0 \over \r} \Bigr) B_{223} \nonumber \\
  & &\hspace{20pt}+{F_a^2 \O_R^2 \over 2a_2^2} (3a_2^2
  A_2 +a_3^2 A_3 -8a_2^4 A_{22} -a_1^2 a_2^4 A_{122} +5a_2^6 A_{222})
  \nonumber \\
  & &\hspace{20pt}-{F_a^4 \O_R^4 \over 8\pi \r} +{2a_1^2 \over a_1^2
    +a_2^2} {\O_R \over \pi \r} r \biggr], \\
  \g_{23} &=& \pi \r\biggl[ -{P_0 \over \r}
  \Bigl( {A_2 \over a_3^2} +{A_3 \over a_2^2} \Bigr)
  +\pi \r A_2 A_3 \nonumber \\ 
  & &\hspace{20pt} +\Bigl( 2\pi \r a_2^2
  A_2 +{3P_0 \over \r} \Bigr) \Bigl( -a_2^2 A_{223} +{1 \over 2}
  B_{223} \Bigr) 
  +{1 \over 2} \Bigl( 2\pi \r a_1^2 A_1 +{3P_0 \over
    \r} \Bigr) B_{123} \nonumber \\
  & &\hspace{20pt}+\Bigl( 2\pi \r a_3^2 A_3 +{3P_0 \over \r} \Bigr) 
  \Bigl( -a_3^2 A_{233} +{1 \over 2} B_{233} \Bigr)\nonumber \\
  & &\hspace{20pt}+{F_a^2 \O_R^2 \over 2} (4A_3 -a_1^2 A_{13} -9a_2^2
  A_{23} +6a_2^4 A_{223}) +{2a_1^2 \over a_1^2 +a_2^2} {\O_R \over \pi\r} s 
\biggr], \\
  \g_{33} &=& \pi \r\biggl[ -A_3 {P_0 \over \r a_3^2} +{1 \over 2} \pi
  \r A_3^2 +\Bigl( 2\pi \r a_3^2 A_3 +{3P_0 \over \r} \Bigr) \Bigl(
  -a_3^2 A_{333} +{1 \over 4} B_{333} \Bigr) \nonumber \\
  & &\hspace{20pt}+{1 \over 4} \Bigl( 2\pi \r a_1^2 A_1 +{3P_0 \over \r}
  \Bigr) B_{133} +{1 \over 4} \Bigl( 2\pi \r a_2^2 A_2 +{3P_0 \over \r}
  \Bigr) B_{233} \nonumber \\
  & &\hspace{20pt}-{F_a^2 \O_R^2 \over 2} (a_1^2 B_{133} +a_2^2 B_{233})
  \biggr].
\eeqa
$\d U$ denotes the gravitational potential induced by the deformation of
the figure. The explicit form of $\d U$ is given in the \S \ref{boundary}.

\section{Second tensor virial equation}

The angular velocity is determined from the second tensor 
virial equation,\cite{chandra67} which is written as 
\beqa
  \int d^3 x \r x_j {\p \over \p x_i} {\cal G} =0.
\eeqa
Substituting Eq. (\ref{integrated}) into the above equation, we obtain 
\beqa
  0&=& -\d_{ij} (\Pi +\d \Pi) -W_{ij} -\d W_{ij} +F_a (F_a -2) (\O_R^2
  +\d \O^2) \d_{1i} (I_{1j} +\d I_{1j}) \nonumber \\
  & &+F_a (F_a +2) (\O_R^2 +\d \O^2)
  \d_{2i} (I_{2j} +\d I_{2j}) \nonumber \\
  & &+{1 \over c^2} \Bigl[ 2\g_i I_{ij}
  +4\g_{11} I_{111j} \d_{1i} +2\g_{12} (I_{122j} \d_{1i} +I_{112j}
  \d_{2i}) \nonumber \\
  & &\hspace{25pt} +2\g_{13} (I_{133j} \d_{i1} +I_{113j} \d_{3i}) +4\g_{22}
  I_{222j} \d_{2i} \nonumber \\
  & &\hspace{25pt}+2\g_{23} (I_{233j} \d_{2i} +I_{223j} \d_{3i})
  +4\g_{33} I_{333j} \d_{3i} \Bigr], \label{totalSTVeq}
\eeqa
where
\beqa
  \Pi&=& \int d^3 x P, \\
  W_{ij} &=&\int d^3 x \r x_i {\p U \over \p x_j} =-2\pi \r A_i I_{ij},
  \\
  I_{ij} &=&\int d^3 x \r x_i x_j = {M \over 5} a_i^2 \d_{ij}, \\
  I_{iijj} &=&\int d^3 x \r x_i^2 x_j^2 ={M \over 35} a_i^2 a_j^2
(1 + 2 \delta_{ij}), \\
  \d \Pi &=& \d \int d^3 x P, \label{deltaPi} \\
  \d W_{ij} &=& \d \int d^3 x \r x_i {\p U \over \p x_j}, \label{deltaW} 
  \\
  \d I_{ij} &=& \d \int d^3 x \r x_i x_j. \label{deltaI}
\eeqa
$M$ denotes the Newtonian mass:
\beqa
  M \equiv \int d^3 x \r.
\eeqa
$\delta \Pi$, $\delta W_{ij}$, and $\delta I_{ij}$ must
be taken into account 
because the ellipsoidal figure deforms 
due to the 1PN effect. The explicit forms are given later.

\subsection{Second virial equations at Newtonian order}

First, we derive equations for the angular velocity and 
axial ratios of the ellipsoid at Newtonian order. 
The second tensor virial equation at Newtonian order is written
\beqa
  0&=& -\d_{ij} \Pi -W_{ij} +F_a (F_a -2) \O_R^2
  \d_{1i} I_{1j} +F_a (F_a +2) \O_R^2 \d_{2i} I_{2j}. 
\eeqa
The non-vanishing components are
\beqa
  \Pi&=& -W_{11} +F_a (F_a -2) \O_R^2 I_{11}, \\
  \Pi&=& -W_{22} +F_a (F_a +2) \O_R^2 I_{22}, \\
  \Pi&=& -W_{33}.
\eeqa
Then, we obtain
\beqa
  & &\O_R^2 \Bigl[ 1+ \Bigl( {2a_1 a_2 \over a_1^2 +a_2^2} \Bigr)^2 \Bigr]
  = 2 \pi \r B_{12}, \label{eqfig1} \\
  & &-{2a_1^2 a_2^2 \over a_1^2 +a_2^2} \O_R^2 = \pi \r (a_1^2 a_2^2
  A_{12} -a_3^2 A_3) . \label{eqfig2}
\eeqa
{}From these equations, we can determine the axial ratios $a_2/a_1$ and
$a_3/a_1$, and the angular velocity.

\subsection{Deformation of the figure}

The deformation of an incompressible 
ellipsoid due to the 1PN gravity is written
by using the Lagrangian displacement vectors $\xi^{(ij)}_k$ 
of $\sum_k \pa_k \xi_k^{(ij)}=0$\cite{chandra67} as
\beqa
  \xi_k ={1 \over c^2} \sum_{ij} S_{ij} \xi_k^{(ij)},
\eeqa
where
\beqa
  \xi_k^{(11)} &=& (x_1, ~0, ~-x_3), \\
  \xi_k^{(12)} &=& (0, ~x_2, ~-x_3), \\
  \xi_k^{(31)} &=& \Bigl( {1 \over 3} x_1^3, ~-x_1^2 x_2, ~0 \Bigr), \\
  \xi_k^{(32)} &=& \Bigl( 0, ~{1 \over 3} x_2^3, ~-x_2^2 x_3 \Bigr), \\
  \xi_k^{(33)} &=& \Bigl( -x_3^2 x_1, ~0, ~{1 \over 3} x_3^3 \Bigr).
\eeqa
Here, we consider only up to the cubic deformation and also assume
triplane symmetries, so that the Lagrangian displacement vectors of
higher order functions in $x_i$ and of even functions can be neglected.

Using the Lagrangian displacement vectors, 
we can calculate Eqs. (\ref{deltaPi}) $\sim$ (\ref{deltaI}).
We find\cite{chandra69}
\beqa
  \d \Pi &=&0, \\
  \d I_{ij} &=&\int d^3 x \r \sum_l \xi_l {\p \over \p x_l} (x_i x_j), \\
  \d I_{11} &=& {2 \over c^2} \Bigl( S_{11} I_{11} +{1 \over 3} S_{31}
  I_{1111} -S_{33} I_{1133} \Bigr), \\
  \d I_{22} &=& {2 \over c^2} \Bigl( S_{12} I_{22} -S_{31} I_{1122} +{1
    \over 3} S_{32} I_{2222} \Bigr), \\
  \d I_{33} &=& {2 \over c^2} \Bigl( -S_{11} I_{33} -S_{12} I_{33}
  -S_{32} I_{2233} +{1 \over 3} S_{33} I_{3333} \Bigr), \\
  \d W_{ij} &=& \pi \r \bigl( -2B_{ij} \d I_{ij} +a_i^2 \d_{ij} \sum_l
  A_{il} \d I_{ll} \bigr), \\
  \d W_{11} &=& {2\pi \r \over c^2} \biggl[ S_{11} \Bigl\{ (a_1^2 A_{11}
  -2B_{11}) I_{11} -a_1^2 A_{13} I_{33} \Bigr\} \nonumber \\
 & &\hspace{25pt} +S_{12} (a_1^2 A_{12}
  I_{22} -a_1^2 A_{13} I_{33}) \nonumber \\
  & &\hspace{25pt}+S_{31} \Bigl\{ {1 \over 3} (a_1^2 A_{11}
  -2B_{11}) I_{1111} -a_1^2 A_{12} I_{1122} \Bigr\} \nonumber \\
  & &\hspace{25pt}+S_{32} \Bigl( {1
    \over 3} a_1^2 A_{12} I_{2222} -a_1^2 A_{13} I_{2233} \Bigr)
  \nonumber \\
  & &\hspace{25pt}+S_{33} 
  \Bigl\{ -(a_1^2 A_{11} -2B_{11}) I_{1133} +{1 \over 3} a_1^2 A_{13}
  I_{3333} \Bigr\} \biggr], \\
  \d W_{22} &=& {2\pi \r \over c^2} \biggl[ S_{11} (a_2^2 A_{12} I_{11}
  -a_2^2 A_{23} I_{33}) \nonumber \\
  & &\hspace{25pt} +S_{12} \Bigl\{ (a_2^2 A_{22} -2B_{22}) I_{22}
  -a_2^2 A_{23} I_{33} \Bigr\} \nonumber \\
  & &\hspace{25pt} +S_{31} \Bigl\{ {1 \over 3} a_2^2 A_{12}
  I_{1111} -(a_2^2 A_{22} -2B_{22}) I_{1122} \Bigr\} \nonumber \\
  & &\hspace{25pt} +S_{32} \Bigl\{ {1
    \over 3} (a_2^2 A_{22} -2B_{22}) I_{2222} -a_2^2 A_{23} I_{2233}
  \Bigr\} \nonumber \\
  & &\hspace{25pt}+S_{33} \Bigl( -a_2^2 A_{12} I_{1133} +{1 \over 3} a_2^2
  A_{23} I_{3333} \Bigr) \biggr], \\
  \d W_{33} &=& {2\pi \r \over c^2} \biggl[ S_{11} \Bigl\{ a_3^2 A_{13}
  I_{11} -(a_3^2 A_{33} -2B_{33}) I_{33} \Bigr\} \nonumber \\
  & &\hspace{25pt} +S_{12} \Bigl\{ a_3^2
  A_{23} I_{22} -(a_3^2 A_{33} -2B_{33}) I_{33} \Bigr\} \nonumber \\
  & &\hspace{25pt}+S_{31} \Bigl( {1 \over 
    3} a_3^2 A_{13} I_{1111} -a_3^2 A_{23} I_{1122} \Bigr) \nonumber \\
  & &\hspace{25pt} +S_{32}
  \Bigl\{ {1 \over 3} a_3^2 A_{23} I_{2222} -(a_3^2 A_{33} -2B_{33})
  I_{2233} \Bigr\} \nonumber \\
  & &\hspace{25pt}+S_{33} \Bigl\{ -a_3^2 A_{13} I_{1133} +{1 \over 3}
  (a_3^2 A_{33} -2B_{33}) I_{3333} \Bigr\} \biggr]. 
\eeqa

\subsection{Second virial equations in the 1PN approximation}

Substituting the 1PN terms into Eq. (\ref{totalSTVeq}), we have the
second virial equations in the 1PN approximation. 
Their components are given as follows:
\begin{enumerate}
\item $i=j=1$:
\beqa
  0&=& 2\g_1 a_1^2 +{12 \over 7} \g_{11} a_1^4 +{2
    \over 7} \g_{12} a_1^2 a_2^2 +{2 \over 7} \g_{13} a_1^2 a_3^2 +F_a
  (F_a -2) a_1^2 \d \O^2 \nonumber \\
  & &+2F_a (F_a -2) \O_R^2 \Bigl( S_{11} a_1^2 +{1 \over 7} S_{31} a_1^4
  -{1 \over 7} S_{33} a_1^2 a_3^2 \Bigr) \nonumber \\
  & &-2\pi \r \Bigl[ S_{11} (a_1^4 A_{11} -2a_1^2 B_{11}
  -a_1^2 a_3^2 A_{13}) +S_{12} (a_1^2 a_2^2 A_{12} -a_1^2 a_3^2 A_{13})
\nonumber \\
  & &\hspace{30pt}
  +{S_{31} \over 7} (a_1^6 A_{11} -2a_1^4 B_{11} -a_1^4 a_2^2 A_{12})
  +{S_{32} \over 7} (a_1^2 a_2^4 A_{12} -a_1^2 a_2^2
  a_3^2 A_{13}) \nonumber \\
  & &\hspace{30pt}+{S_{33} \over 7} (-a_1^4 a_3^2 A_{11} +2a_1^2 a_3^2
  B_{11} +a_1^2 a_3^4 A_{13}) \Bigr],
\eeqa
\item $i=j=2$:
\beqa
  0&=& 2\g_2 a_2^2 +{2 \over 7} \g_{12} a_1^2 a_2^2
  +{12 \over 7} \g_{22} a_2^4 +{2 \over 7} \g_{23} a_2^2 a_3^2 +F_a
  (F_a +2) a_2^2 \d \O^2 \nonumber \\
  & &+2F_a (F_a +2) \O_R^2 \Bigl( S_{12} a_2^2 -{1 \over 7} S_{31} a_1^2
  a_2^2 +{1 \over 7} S_{32} a_2^4 \Bigr) \nonumber \\
  & &-2\pi \r \Bigl[ S_{11} (a_1^2 a_2^2 A_{12} -a_2^2
  a_3^2 A_{23}) +S_{12} (a_2^4 A_{22} -2a_2^2 B_{22} -a_2^2 a_3^2
  A_{23}) \nonumber \\
  & &\hspace{30pt}+{S_{32} \over 7} (a_2^6 A_{22} -2a_2^4 B_{22}
  -a_2^4 a_3^2 A_{23}) 
  +{S_{33} \over 7} (-a_1^2 a_2^2 a_3^2 A_{12} +a_2^2
  a_3^4 A_{23}) \nonumber \\
  & &\hspace{30pt}+{S_{31} \over 7} (a_1^4 a_2^2 A_{12} -a_1^2 a_2^4
  A_{22} +2a_1^2 a_2^2 B_{22}) \Bigr],
\eeqa
\item $i=j=3$:
\beqa
  0&=& 2\g_3 a_3^2 +{2 \over 7} \g_{13} a_1^2 a_3^2
  +{2 \over 7} \g_{23} a_2^2 a_3^2 +{12 \over 7} \g_{33} a_3^4 \nonumber \\
  & &-2\pi \r \Bigl[ S_{11} (a_1^2 a_3^2 A_{13} -a_3^4
  A_{33} +2a_3^2 B_{33}) +S_{12} (a_2^2 a_3^2 A_{23} -a_3^4 A_{33}
  +2a_3^2 B_{33}) \nonumber \\
  & &\hspace{30pt}+{S_{33} \over 7} (a_3^6 A_{33} -2a_3^4 B_{33}
  -a_1^2 a_3^4 A_{13}) +{S_{31} \over 7} (a_1^4 a_3^2 A_{13} -a_1^2 a_2^2
  a_3^2 A_{23}) \nonumber \\
  & &\hspace{30pt}+{S_{32} \over 7} (a_2^4 a_3^2 A_{23} -a_2^2 a_3^4
  A_{33} +2a_2^2 a_3^2 B_{33}) \Bigr].
\eeqa
\end{enumerate}

Subtracting the $(2,2)$ component of the 1PN second virial equation from
the $(1,1)$ component, we have the equation for $\d \O^2$ as
\beqa
  0&=& 2\g_1 a_1^2 -2\g_2 a_2^2 +{2 \over 7} (6
  \g_{11} a_1^4 +\g_{13} a_1^2 a_3^2 -6\g_{22} a_2^4 -\g_{23} a_2^2
  a_3^2) \nonumber \\
 & & -(a_1^2 -a_2^2) \Bigl[ 1+ \Bigl( {2a_1 a_2 \over a_1^2 +a_2^2}
  \Bigr)^2 \Bigr] \d \O^2  \nonumber \\
  & &-2\pi \r \Bigl[ S_{11} \tau_1 +S_{12} \tau_2 +{a_1^2 \over 7}
  S_{31} (\tau_1 -\tau_2) +{a_2^2 \over 7} S_{32} \tau_2 -{a_3^2 \over
    7} S_{33} \tau_1 \Bigr],~~ \label{do1}
\eeqa
where we use the equilibrium equations of Newtonian order and define
new variables as
\beqa
  \tau_1 &\equiv& 3a_1^4 A_{11} -a_1^2 a_2^2 A_{12} -a_1^2 a_3^2 A_{13}
  +a_2^2 a_3^2 A_{23} -2a_3^2 A_3, \\
  \tau_2 &\equiv& a_1^2 a_2^2 A_{12} -3a_2^4 A_{22} -a_1^2 a_3^2 A_{13}
  +a_2^2 a_3^2 A_{23} +2a_3^2 A_3.
\eeqa
Then, $\d \O^2$ is expressed as
\beqa
  \d \O^2 &=& {\O_R^2 \over \pi \r (a_1^2 -a_2^2) B_{12}} \Bigl[ \g_1
  a_1^2 - \g_2 a_2^2 +{1 \over 7} (6 \g_{11} a_1^4 +\g_{13} a_1^2 a_3^2
  -6\g_{22} a_2^4 -\g_{23} a_2^2 a_3^2) \nonumber \\
  & &\hspace{10pt}-\pi \r \Bigl( S_{11} \tau_1 +S_{12} \tau_2 +{a_1^2
    \over 7} S_{31} (\tau_1 -\tau_2) +{a_2^2 \over 7} S_{32} \tau_2
  -{a_3^2 \over 7} S_{33} \tau_1 \Bigr) \Bigr]. \label{deltaomega}
\eeqa
By calculating $(1,1)-(3,3)$ and $(2,2)-(3,3)$, we can find
other expressions of $\d \O^2$  as 
\beqa
  0&=& 2\g_1 a_1^2 -2\g_3 a_3^2 +{2 \over 7} (6
  \g_{11} a_1^4 +\g_{12} a_1^2 a_2^2 -\g_{23} a_2^2 a_3^2 -6\g_{33}
  a_3^4) +F_a (F_a -2) a_1^2 \d \O^2  \nonumber \\
  & &-2\pi \r \Bigl[ S_{11} (\tau_1 -\tau_3) -S_{12} \tau_3 +{a_1^2
    \over 7} S_{31} \tau_1 -{a_2^2 \over 7} S_{32} \tau_3 -{a_3^2 \over
    7} S_{33} (\tau_1 -\tau_3) \Bigr], \label{do2} \\
  0&=& 2\g_2 a_2^2 -2\g_3 a_3^2 +{2 \over 7}
  (\g_{12} a_1^2 a_2^2 +6\g_{22} a_2^4 -\g_{13} a_1^2 a_3^2 -6\g_{33}
  a_3^4) +F_a (F_a +2) a_2^2 \d \O^2  \nonumber \\
  & &-2\pi \r \Bigl[ -S_{11} \tau_3 -S_{12} (\tau_2 +\tau_3) +{a_1^2
    \over 7} S_{31} \tau_2 -{a_2^2 \over 7} S_{32} (\tau_2 +\tau_3)
  +{a_3^2 \over 7} S_{33} \tau_3 \Bigr], \label{do3}
\eeqa
where
\beqa
  \tau_3 \equiv a_1^2 a_3^2 A_{13} -3a_3^4 A_{33} -a_1^2 a_2^2 A_{12}
  +a_2^2 a_3^2 A_{23} +2a_3^2 A_3.
\eeqa
Eliminating $\d \O^2$ from Eq. (\ref{do2}) by using Eq. (\ref{do3}), we
obtain the identity 
\beqa
  0 &=&(3a_1^2 +a_2^2) a_2^2 \Bigl[ \g_1 a_1^2 -\g_3 a_3^2 +{1 \over
    7} (6 \g_{11} a_1^4 +\g_{12} a_1^2 a_2^2 -\g_{23} a_2^2 a_3^2
  -6\g_{33} a_3^4) \Bigr] \nonumber \\
  & &+(a_1^2 +3a_2^2) a_1^2 \Bigl[ \g_2 a_2^2 -\g_3 a_3^2 +{1 \over 7}
  (\g_{12} a_1^2 a_2^2 +6\g_{22} a_2^4 -\g_{13} a_1^2 a_3^2 -6\g_{33}
  a_3^4) \Bigr] \nonumber \\
  & &-\pi \r (3a_1^2 +a_2^2) a_2^2 \Bigl[ S_{11} (\tau_1 -\tau_3)
  -S_{12} \tau_3 +{a_1^2 \over 7} S_{31} \tau_1 \nonumber \\
  & & \hspace{80pt}-{a_2^2 \over 7} S_{32}
  \tau_3 -{a_3^2 \over 7} S_{33} (\tau_1 -\tau_3) \Bigr]\nonumber \\
  & &-\pi \r (a_1^2 +3a_2^2) a_1^2 \Bigl[ -S_{11} \tau_3 -S_{12}
  (\tau_2 +\tau_3) +{a_1^2 \over 7} S_{31} \tau_2 \nonumber \\
  & & \hspace{80pt} -{a_2^2 \over 7}
  S_{32} (\tau_2 +\tau_3) +{a_3^2 \over 7} S_{33} \tau_3
  \Bigr]. \label{cond10}
\eeqa
Equation (\ref{cond10}) is one of the equations to determine $p$, $q$,
$r$, $s$, and $S_{ij}$.

\section{Boundary conditions}\label{boundary}

In this section we derive equations to determine the coefficients of the
 1PN velocity potential and the deformation of the figure, 
 $p$, $q$, $r$, $s$, and $S_{ij}$, from the 
boundary conditions on the stellar surface. 
The stellar surfaces of the Riemann ellipsoid and its deformed figure are 
expressed as $S_R(x)=0$ and $S(x)=0$, respectively, where 
\beqa
  S_R(x) &=& \sum_l {x_l^2 \over a_l^2} -1, \\
  S(x) &=& \sum_l {x_l^2 \over a_l^2} -1 -\sum_j 
\xi_j {\p S_R \over \p x_j} \nonumber   \\
  &=& S_R(x) -{2 \over c^2} \biggl[ S_{11} \Bigl(
  {x_1^2 \over a_1^2} -{x_3^2 \over a_3^2} \Bigr) +S_{12} \Bigl( {x_2^2
    \over a_2^2} -{x_3^2 \over a_3^2} \Bigr) +S_{31} \Bigl( {x_1^4 \over 
    3a_1^2} -{x_1^2 x_2^2 \over a_2^2} \Bigr) \nonumber \\
  & &\hspace{70pt}+S_{32} \Bigl( {x_2^4
    \over 3a_2^2} -{x_2^2 x_3^2 \over a_3^2} \Bigr) +S_{33} \Bigl(
  {x_3^4 \over 3a_3^2} -{x_3^2 x_1^2 \over a_1^2} \Bigr) \biggr].
\eeqa
The boundary conditions for the continuity equation and 
integrated Euler equation are, respectively, 
\beqa
  &(A)&~~ \sum_j C_j {\p S \over \p x_j} =0~~~~{\rm on}~~S(x)=0,
  \label{velocitycond} \\
  &(B)&~~ {P \over \r} =0~~~~~{\rm on}~~S(x)=0. \label{pressurecond}
\eeqa
The condition (A) comes from the constraint that the
normal component of the velocity on the surface vanishes. 
The condition (B) determines the stellar surface. 
Eqs. (\ref{velocitycond}) and (\ref{pressurecond}) must hold 
to order of $c^{-2}$. 

\subsection{Condition (A)}

Substituting $C_i$ given by Eq. (\ref{1PNvelocity}) into 
Eq. (\ref{velocitycond}), we obtain an equation of 1PN order as 
\beq
0=x_1x_2(c_0 + c_1 x_1^2 +c_2 x_2^2 + c_3 x_3^2),\label{c0123}
\eeq
where $c_k~(k=0,1,2,3)$ are functions of $a_k$, $p$, $q$, $r$, $s$, and
$S_{ij}$. Equation (\ref{c0123}) must be satisfied on $S_R$. 
Then, we
obtain the following three equations for $p$, $q$, $r$, $s$ and $S_{ij}$:
\beqa
C^{(A)}_1\equiv
  & &{a_1^2 +a_2^2 \over a_1^4 a_2^2} p +{a_1^2 +3a_2^2 \over a_1^2
    a_2^2} q +{4\O \over a_1^2 +a_2^2} \Bigl[ {1 \over a_1^2} 
    (-S_{11} +S_{12})-{5 \over 3} S_{31} \Bigr] \nonumber \\
  & &\hspace{10pt}-3\pi \r \O \Bigl( {A_0
    \over a_1^2} -A_1 \Bigr) {a_1^2 -a_2^2 \over a_1^2 a_2^2} -{1 \over
    2} F_a^2 \O^3 {a_1^2 -a_2^2 \over a_1^2 a_2^2} \nonumber \\
  & &\hspace{10pt}+{\pi \r F_a \O \over 
    2a_1^2} \Bigl[ \Bigl( 7{a_1^2 \over a_2^2} -1 \Bigr) B_{11} +\Bigl(
  7{a_2^2 \over a_1^2} -1 \Bigr) B_{12} +4(a_1^2 +a_2^2) B_{112} \Bigr]
  =0,~~ \label{cond2} \\
C^{(A)}_2\equiv
  & &{a_1^2 +a_2^2 \over a_1^2 a_2^4} p +{3a_1^2 +a_2^2 \over a_1^2
    a_2^2} r +{4\O \over a_1^2 +a_2^2} \Bigl[ {1 \over a_2^2} 
    (-S_{11} +S_{12})+{a_1^2 \over a_2^2} S_{31}+{2 \over 3} 
    S_{32} \Bigr] \nonumber \\
  & &\hspace{10pt} 
  -3\pi \r \O \Bigl( {A_0
    \over a_2^2} -A_2 \Bigr) {a_1^2 -a_2^2 \over a_1^2 a_2^2} -{1 \over
    2} F_a^2 \O^3 {a_1^2 -a_2^2 \over a_1^2 a_2^2} \nonumber \\
  & &\hspace{10pt}+{\pi \r F_a \O \over 
    2a_2^2} \Bigl[ \Bigl( 7{a_1^2 \over a_2^2} -1 \Bigr) B_{12} +\Bigl(
  7{a_2^2 \over a_1^2} -1 \Bigr) B_{22} +4(a_1^2 +a_2^2) B_{122} \Bigr]
  =0,~~ \label{cond3} \\
C^{(A)}_3\equiv
  & &{a_1^2 +a_2^2 \over a_1^2 a_2^2 a_3^2} p +\Bigl( {1 \over a_1^2}
  +{1 \over a_2^2} +{2 \over a_3^2} \Bigr) s
  +{4 \O \over a_1^2 +a_2^2} \Bigl[ {1 \over a_3^2} (-S_{11} +S_{12}) 
  -{a_2^2 \over a_3^2} S_{32}+S_{33} \Bigr] \nonumber \\
  & &\hspace{10pt}
  -3\pi \r \O \Bigl( {A_0
    \over a_3^2} -A_3 \Bigr) {a_1^2 -a_2^2 \over a_1^2 a_2^2} \nonumber \\
  & &\hspace{10pt}+{\pi \r F_a \O \over 
    2a_3^2} \Bigl[ \Bigl( 7{a_1^2 \over a_2^2} -1 \Bigr) B_{13} +\Bigl(
  7{a_2^2 \over a_1^2} -1 \Bigr) B_{23} +4(a_1^2 +a_2^2) B_{123} \Bigr]
  =0.~~ \label{cond4} 
\eeqa
We can make two equations which do not contain $S_{11}$, $S_{12}$ and
  $p$ as
\beqn
&&C^{(A)}_1a_1^2-C^{(A)}_2a_2^2=0,\label{conda1} \\
&&C^{(A)}_1a_1^2-C^{(A)}_3a_3^2=0.\label{conda2}
\eeqn
Thus, Eqs. (\ref{conda1}) and (\ref{conda2}) depend only on 
the coefficients $q$, $r$, $s$, $S_{31}$, $S_{32}$ and $S_{33}$; i.e., 
the coefficients for biquadratic deformation. 
This is an important feature in determining these coefficients. 

\subsection{Condition (B)}

To obtain ${\cal G}$ consistently, we must calculate the correction of 
the self-gravity potential due to the deformation of the figure $\d U$, 
which is given by\cite{chandra67}
\beqa
  \d U = {1 \over c^2} \sum_{ij} S_{ij} \d U^{(ij)}, 
\eeqa
where 
\beqa
  \d U^{(ij)} =-\sum_k {\p \over \p x_k} \int {\r (x') \xi_k^{(ij)} (x') \over
    |x -x'|} d^3 x'. \label{deltaU}
\eeqa
Substituting the Lagrangian displacement vectors 
into Eq. (\ref{deltaU}), we have
\beqa
  \d U^{(11)} &=& -D_{1,1} +D_{3,3}, \\
  \d U^{(12)} &=& -D_{2,2} +D_{3,3}, \\
  \d U^{(31)} &=& -{1 \over 3} D_{111,1} +D_{112,2}, \\
  \d U^{(32)} &=& -{1 \over 3} D_{222,2} +D_{223,3}, \\
  \d U^{(33)} &=& -{1 \over 3} D_{333,3} +D_{133,1},
\eeqa
where \cite{chandra69} for $i \neq j$, 
\beqa
  D_{iii} &=&\pi \r \Bigl[ a_i^6 \bigl( A_{iii} -\sum_l A_{iiil} x_l^2
  \bigr) x_i^3 \nonumber \\
  & &\hspace{20pt}+{3 \over 4} a_i^4 \bigl( B_{ii} -2\sum_l B_{iil} x_l^2
  +\sum_l \sum_m B_{iilm} x_l^2 x_m^2 \bigr) x_i \Bigr], \\
  D_{iij} &=&\pi \r \Bigl[ a_i^4 a_j^2 \bigl( A_{iij} -\sum_l A_{iijl}
  x_l^2 \bigr) x_i^2 x_j \nonumber \\
  & &\hspace{20pt}+{1 \over 4} a_i^2 a_j^2 \bigl( B_{ij} -2\sum_l 
  B_{ijl} x_l^2 +\sum_l \sum_m B_{ijlm} x_l^2 x_m^2 \bigr) x_j \Bigr].
\eeqa
Substituting $\d U$ and the equation $S=0$ into Eq. (\ref{integrated}),
we obtain 
\beqa
  \Bigl( {P \over \r} \Bigr)_S &=& -{2P_0 \over \r c^2} \biggl[ S_{11}
  \Bigl( {x_1^2 \over a_1^2} -{x_3^2 \over a_3^2} \Bigr) +S_{12} \Bigl(
  {x_2^2 \over a_2^2} -{x_3^2 \over a_3^2} \Bigr) +S_{31} \Bigl( {x_1^4
    \over 3a_1^2} -{x_1^2 x_2^2 \over a_2^2} \Bigr) \nonumber \\
  & &\hspace{30pt}+S_{32} \Bigl(
  {x_2^4 \over 3a_2^2} -{x_2^2 x_3^2 \over a_2^2} \Bigr) +S_{33} \Bigl(
  {x_3^4 \over 3a_3^2} -{x_1^2 x_3^2 \over a_1^2} \Bigr) \biggr]
  \nonumber \\
  & &-{1 \over c^2} \biggl[ S_{11} (D_{1,1} -D_{3,3}) +S_{12} (D_{2,2}
  -D_{3,3}) +S_{31} \Bigl( {1 \over 3} D_{111,1} -D_{112,2} \Bigr)
  \nonumber \\
  & &\hspace{30pt}+S_{32} \Bigl( {1 \over 3} D_{222,2} -D_{223,3} \Bigr)
  +S_{33} \Bigl( -D_{133,1} +{1 \over 3} D_{333,3} \Bigr) \biggr]
  \nonumber \\
  & &-{1 \over c^2} \d \O^2 \biggl[ {F_a^2 \over 2} (x_1^2 +x_2^2) +F_a
  (x_2^2 -x_1^2) \biggr] \nonumber \\
  & & -{1 \over c^2} \bigl( \g_0 +\sum_l \g_l x_l^2
  +\sum_{l \le m} \g_{lm} x_l^2 x_m^2 \bigr) +{\rm const}\nonumber  \\
  &\equiv& {1 \over c^2} \Bigl[ Q_1 x_1^2 +Q_2 x_2^2 +Q_3 x_3^2 +Q_{11}
  x_1^4 +Q_{22} x_2^4 +Q_{33} x_3^4 \nonumber \\
  & &\hspace{20pt} +Q_{12} x_1^2 x_2^2 +Q_{13} x_1^2
  x_3^2 +Q_{23} x_2^2 x_3^2 +{\rm const} \Bigr] , 
\eeqa
where
\beqa
  Q_1 &=& -\g_1 -{F_a \over 2} (F_a -2) \d \O^2 -S_{11} \pi \r \Bigl(
  {2a_3^2 \over a_1^2} A_3 -3a_1^2 A_{11} +a_3^2 A_{13} \Bigr) \nonumber \\
  & &+S_{12}
  \pi \r (a_2^2 A_{12} -a_3^2 A_{13}) \nonumber \\
  & &-S_{31} \pi \r a_1^2 \Bigl( a_1^4
  A_{111} -a_1^2 a_2^2 A_{112} -{3 \over 2} a_1^2 B_{111} +{1 \over 2}
  a_2^2 B_{112} \Bigr) \nonumber \\
  & & +S_{32} \pi \r {a_2^2 \over 2} (a_2^2 B_{122}
  -a_3^2 B_{123}) 
  +S_{33} \pi \r {a_3^2 \over 2} (a_3^2 B_{133} -3a_1^2 B_{113}), \\
  Q_2 &=& -\g_2 -{F_a \over 2} (F_a +2) \d \O^2 +S_{11} \pi \r (a_1^2
  A_{12} -a_3^2 A_{23}) \nonumber \\
  & & -S_{12} \pi \r \Bigl( {2a_3^2 \over a_2^2} A_3
  -3a_2^2 A_{22} +a_3^2 A_{23} \Bigr) 
  +S_{31} \pi \r {a_1^2 \over 2}
  (a_1^2 B_{112} -3a_2^2 B_{122}) \nonumber \\
  & &-S_{32} \pi \r a_2^2 \Bigl(a_2^4
  A_{222} -a_2^2 a_3^2 A_{223} -{3 \over 2} a_2^2 B_{222} +{1 \over 2}
  a_3^2 B_{223} \Bigr) \nonumber \\
  & &+S_{33} \pi \r {a_3^2 \over 2} (a_3^2 B_{233} -a_1^2 B_{123}), \\
  Q_3 &=& -\g_3 +S_{11} \pi \r (2A_3 +a_1^2 A_{13} -3a_3^2 A_{33})
  +S_{12} \pi \r (2A_3 +a_2^2 A_{23} -3a_3^2 A_{33}) \nonumber \\
  & &+S_{31} \pi \r
  {a_1^2 \over 2} (a_1^2 B_{113} -a_2^2 B_{123}) +S_{32} \pi \r {a_2^2
    \over 2} (a_2^2 B_{223} -3a_3^2 B_{233}) \nonumber \\
  & &-S_{33} \pi \r a_3^2 \Bigl(
  a_3^4 A_{333} -a_1^2 a_3^2 A_{133} -{3 \over 2} a_3^2 B_{333} +{1
    \over 2} a_1^2 B_{133} \Bigr), \\
  Q_{11} &=& -\g_{11} -S_{31} \pi \r a_1^2 \Bigr( {2a_3^2 \over 3a_1^4}
  A_3 -{5 \over 3} a_1^4 A_{1111} +a_1^2 a_2^2 A_{1112} +{5 \over 4}
  a_1^2 B_{1111} -{1 \over 4} a_2^2 B_{1112} \Bigr) \nonumber \\
  & &-S_{32} \pi \r
  {a_2^2 \over 4} (a_2^2 B_{1122} -a_3^2 B_{1123}) -S_{33} \pi \r {a_3^2 
    \over 4} (a_3^2 B_{1133} -5a_1^2 B_{1113}), \\
  Q_{22} &=& -\g_{22} -S_{31} \pi \r {a_1^2 \over 4} (a_1^2 B_{1122}
  -5a_2^2 B_{1222}) \nonumber \\
  & & -S_{32} \pi \r a_2^2 \Bigl( {2a_3^2 \over 3a_2^4}
  A_3 -{5 \over 3} a_2^4 A_{2222} +a_2^2 a_3^2 A_{2223} +{5 \over 4}
  a_2^2 B_{2222} -{1 \over 4} a_3^2 B_{2223} \Bigr) \nonumber \\
  & &-S_{33} \pi \r {a_3^2 \over 4} (a_3^2 B_{2233} -a_1^2 B_{1223}), \\ 
  Q_{33} &=& -\g_{33} -S_{31} \pi \r {a_1^2 \over 4} (a_1^2 B_{1133}
  -a_2^2 B_{1233}) -S_{32} \pi \r {a_2^2 \over 4} (a_2^2 B_{2233}
  -5a_3^2 B_{2333}) \nonumber \\
  & &-S_{33} \pi \r a_3^2 \Bigl( {2 \over 3a_3^2} A_3 -{5 
  \over 3} a_3^4 A_{3333} +a_1^2 a_3^2 A_{1333} +{5 \over 4} a_3^2
  B_{3333} -{1 \over 4} a_1^2 B_{1333} \Bigr), \nonumber \\
  && \\
  Q_{12} &=& -\g_{12} +S_{31} \pi \r a_1^2 \Bigl( {2 a_3^2 \over a_1^2
    a_2^2} A_3 +a_1^4 A_{1112} -3a_1^2 a_2^2 A_{1122} -{3 \over 2} a_1^2
  B_{1112} +{3 \over 2} a_2^2 B_{1122} \Bigr) \nonumber \\
  & &+S_{32} \pi \r a_2^2 \Bigl( a_2^4 A_{1222} -a_2^2 a_3^2 A_{1223}
  -{3 \over 2} a_2^2 B_{1222} +{1 \over 2} a_3^2 B_{1223} \Bigr) \nonumber \\
  & & -S_{33}
  \pi \r {a_3^2 \over 2} \Bigl( a_3^2 B_{1233} -3a_1^2 B_{1123} \Bigr), \\
  Q_{13} &=& -\g_{13} +S_{31} \pi \r a_1^2 \Bigl( a_1^4 A_{1113} -a_1^2
  a_2^2 A_{1123} -{3 \over 2} a_1^2 B_{1113} +{1 \over 2} a_2^2 B_{1123}
  \Bigr) \nonumber \\
  & & -S_{32} \pi \r {a_2^2 \over 2} \Bigl( a_2^2 B_{1223} -3a_3^2
  B_{1233} \Bigr) \nonumber \\
  & &+S_{33} \pi \r a_3^2 \Bigl( {2 \over a_1^2} A_3 +a_3^4 A_{1333} 
  -3a_1^2 a_3^2 A_{1133} -{3 \over 2} a_3^2 B_{1333} +{3 \over 2} a_1^2
  B_{1133} \Bigr), \\
  Q_{23} &=& -\g_{23} -S_{31} \pi \r {a_1^2 \over 2} \Bigl( a_1^2
  B_{1123} -3a_2^2 B_{1223} \Bigr) \nonumber \\
  & & +S_{32} \pi \r a_2^2 \Bigl( {2 \over
    a_2^2} A_3 +a_2^4 A_{2223} -3a_2^2 a_3^2 A_{2233} -{3 \over 2}a_2^2
  B_{2223} +{3 \over 2}a_3^2 B_{2233} \Bigr) \nonumber \\
  & &+S_{33} \pi \r a_3^2 \Bigl( a_3^4 A_{2333}
  -a_1^2 a_3^2 A_{1233} -{3 \over 2} a_3^2 B_{2333} +{1 \over 2} a_1^2
  B_{1233} \Bigr).
\eeqa
Since $(P/\r)_S$ must vanish on the surface, 
we have five conditions,
\beqa
  a_1^4 Q_{11} +a_2^4 Q_{22} -a_1^2 a_2^2 Q_{12} =0, \label{cond5} \\
  a_2^4 Q_{22} +a_3^4 Q_{33} -a_2^2 a_3^2 Q_{23} =0, \label{cond6} \\
  a_3^4 Q_{33} +a_1^4 Q_{11} -a_3^2 a_1^2 Q_{13} =0, \label{cond7} \\
  a_1^4 Q_{11} -a_2^4 Q_{22} +a_1^2 Q_1 -a_2^2 Q_2 =0, \label{cond8} \\
  a_3^4 Q_{33} -a_1^4 Q_{11} +a_3^2 Q_3 -a_1^2 Q_1 =0. \label{cond9} 
\eeqa

When we substitute $\d \O^2$ into Eqs. (\ref{cond8}) and
(\ref{cond9}), we have two identical equations:
\beqa
  & &(a_3^4 Q_{33} +a_1^4 Q_{11} -a_1^2 a_3^2 Q_{13}) -(a_2^4 Q_{22} +a_3^4 
  Q_{33} -a_2^2 a_3^2 Q_{23}) \nonumber \\
  & & =7(a_1^4 Q_{11} -a_2^4 Q_{22} +a_1^2 Q_1
  -a_2^2 Q_2), \\
  & &(a_2^4 Q_{22} +a_3^4 Q_{33} -a_2^2 a_3^2 Q_{23}) -(a_1^4 Q_{11} +a_2^4 
  Q_{22} -a_1^2 a_2^2 Q_{12}) \nonumber \\
  & & =7(a_3^4 Q_{33} -a_1^4 Q_{11} +a_3^2 Q_3
  -a_1^2 Q_1).
\eeqa
This implies that Eqs. (\ref{cond8}) and (\ref{cond9}) can be derived from 
Eqs. (\ref{cond5}) $\sim$ (\ref{cond7}), and only three of 
five conditions (\ref{cond5}) $\sim$ (\ref{cond9}) are available. Here, 
we choose Eqs. (\ref{cond5}) $\sim$ (\ref{cond7}) as the equations 
to be solved. Then, the three equations do not contain 
$p$, $S_{11}$ and $S_{12}$, so that they depend only on $q$, $r$, $s$, 
$S_{31}$, $S_{32}$ and $S_{33}$. This means that if we use 
Eqs. (\ref{cond5}) $\sim$ (\ref{cond7}) together with 
Eqs. (\ref{cond1}), (\ref{conda1}) and (\ref{conda2}), we can determine 
the six coefficients. 

The question arises when we determine the rest three coefficients 
$p$, $S_{11}$ and $S_{12}$ because 
we have only {\it two} equations (\ref{cond10}) 
and (\ref{cond2}) (or (\ref{cond3}) or (\ref{cond4})) 
for the {\it three} variables. There is one degree of freedom. 
This is because there is no unique definition of the 1PN solution 
as the counterpart of a Newtonian solution, as pointed out by Chandrasekhar 
\cite{chandra65,chandra67} and Bardeen.\cite{Bardeen}
(Or, we may say that we have a degree to fix the PN correction of 
the angular velocity ($\delta \Omega$) or axial ratio ($S_{11}$ or $S_{12}$) 
in obtaining a PN equilibrium solution.)
There are many possible ways to compare a 1PN solution with a 
Newtonian solution, and to fix the 1PN counterpart to a Newtonian solution, 
we must impose an additional condition.  
(For example, Bardeen has proposed that 
we should compare two solutions fixing the angular momentum and 
the baryon number.\cite{Bardeen}) 

Since the method for connecting a 1PN solution to a Newtonian solution 
can be arbitrarily chosen, in this paper, we simply give the condition as
\beq
  S_{11} = S_{12}, \label{cond11}
\eeq
i.e., we fix the PN correction of the axial ratio. 
The reason why we choose this condition is simply that 
$S_{11}$ and $S_{12}$ in 
Eqs. (\ref{cond2}) $\sim$ (\ref{cond4}) 
are erased, and manipulations are simplified. 
Then, we have three equations (\ref{cond10}),
(\ref{cond2}), and (\ref{cond11}) 
for solving $p$, $S_{11}$, and $S_{12}$. 
$\delta \Omega$ is determined from Eq. (3.38) if we obtain these 
coefficients. 

\section{The total energy and angular momentum}

\subsection{Total energy}

The total energy is written as
\beqa
  E &=& \int d^3 x \r \biggl[ {v^2 \over 2} -{1 \over 2} U +{1 \over c^2}
  \Bigl( {5 \over 8} v^4 +{5 \over 2} v^2 U
  +{1 \over 2} \hat \b_i v^i +{P
    \over \r} v^2 -{5 \over 2} U^2 \Bigr)
  +O(c^{-4})\biggr] \nonumber \\
  &=& E_{\rm N} +{1 \over c^2} E_{\rm PN}+O(c^{-4}), \label{totenergy}
\eeqa
where
\beqa
  E_{\rm N} &=&{M_{\ast} \over 10} \Bigl[ F_a^2 \O_R^2 (a_{1 \ast}^2
  +a_{2 \ast}^2) -4\pi \r A_{0 \ast} \Bigr], \\
  E_{\rm PN} &=& E_{\rm N \ra PN} +E_{v^4} +E_{v^2 U} +E_{\b_i v^i}
  +E_{Pv^2/\r} +E_{U^2}, \\
  E_{\rm N \ra PN} &=&-E_{\rm N} \Bigl[ {1 \over 6} F_a^2 \O_R^2 
  (a_{1 \ast}^2+a_{2 \ast}^2) +4\pi \r A_{0 \ast} \Bigr] \nonumber \\
  & &+{M_{\ast} \over 5} F_a^2 \O_R^2 \Bigl[ a_{1 \ast}^2
  S_{11} +a_{2 \ast}^2 S_{12} +{1 \over 7} a_{1 \ast}^2 (a_{1 \ast}^2
  -a_{2 \ast}^2) S_{31} +{1 \over 7} a_{2 \ast}^4
  S_{32} \nonumber \\
  &&\hspace{60pt}-{1 \over 7} a_{1 \ast}^2 a_{3 \ast}^2 S_{33} \Bigr]
  \nonumber \\
  & &+{M_{\ast} \over 10} F_a^2 \d
  \O^2 (a_{1 \ast}^2 +a_{2 \ast}^2) \nonumber \\
  & & +{M_{\ast} \over 5} F_a \O_R (a_{1 \ast}^2 +a_{2 \ast}^2) 
   \Bigl( p +{3 \over 7} q a_{1 \ast}^2 +{3 \over 7} r a_{2 \ast}^2 
  +{1 \over 7} s a_{3 \ast}^2 \Bigr)   \nonumber \\
  & &+{2M_{\ast} \over 5} \pi \r \Bigl[ S_{11} (a_{1 \ast}^2 A_{1 \ast}
  -a_{3 \ast}^2 A_{3 \ast}) +S_{12} (a_{2 \ast}^2 A_{2 \ast} -a_{3
  \ast}^2 A_{3 \ast}) \nonumber \\
  & &\hspace{50pt}+{1 \over 7} a_{1 \ast}^2 
  S_{31} (a_{1 \ast}^2 A_{1 \ast} -a_{2 \ast}^2 A_{2 \ast})
  +{1 \over 7} a_{2 \ast}^2 S_{32} (a_{2 \ast}^2 A_{2 \ast} -a_{3
  \ast}^2 A_{3 \ast}) \nonumber \\
  & &\hspace{50pt}-{1 \over 7} a_{3 \ast}^2 S_{33} (a_{1 \ast}^2 A_{1
    \ast} -a_{3 \ast}^2 A_{3 \ast}) \Bigr], \label{ENPN} \\
  E_{v^4} &=&{M_{\ast} \over 56} F_a^4 \O_R^4 (3a_{1 \ast}^4 +2a_{1
    \ast}^2 a_{2 \ast}^2 +3a_{2 \ast}^4), \\
  E_{v^2 U} &=&{M_{\ast} \over 7} \pi \r F_a^2 \O_R^2 \Bigl[ 3A_{0
    \ast} (a_{1 \ast}^2 +a_{2 \ast}^2) 
  -(a_{1 \ast}^4 A_{1 \ast} +a_{2 \ast}^4 A_{2 \ast}) \Bigr], \\
  E_{\b_i v^i} &=&{M_{\ast} \over 35} \pi \r F_a^2 \O_R^2 \Bigl[
  a_{1 \ast}^2 (a_{2 \ast}^2 - 7a_{1 \ast}^2) A_{1 \ast} 
  +a_{2 \ast}^2 (a_{1 \ast}^2 - 7a_{2 \ast}^2) A_{2 \ast} \nonumber \\
  & &~~-2a_{1 \ast}^2 a_{2 \ast}^2 (a_{1 \ast}^2 +a_{2 \ast}^2) 
  A_{12 \ast} \Bigr], \\
  E_{Pv^2/\r} &=&{2M_{\ast} \over 35} {P_0 \over \r} F_a^2 \O_R^2 (a_{1 
    \ast}^2 +a_{2 \ast}^2), \\
  E_{U^2} &=&-{M_{\ast} \over 7} (\pi \r)^2 (11A_{0 \ast}^2 +a_{1
    \ast}^4 A_{1 \ast}^2 +a_{2 \ast}^4 A_{2 \ast}^2 +a_{3 \ast}^4 A_{3
    \ast}^2).
\eeqa
In the above equations, we have used Eq. (\ref{deltaomega}) for $\d
    \O^2$ and the conserved mass
\beqa
  M_{\ast} &\equiv& \int d^3 x \r \Bigl[ 1 +{1 \over c^2} \Bigl( {v^2
      \over 2} +3U \Bigr) \Bigr], \nonumber \\
  &=& M \Bigl[ 1+{1 \over c^2} \Bigl\{ {1 \over 10} F_a^2 \O_R^2 (a_1^2
  +a_2^2) +{12 \over 5} \pi \r A_0 \Bigr\} \Bigr],
\eeqa
instead of the Newtonian mass $M$. Also we substitute the mean radius
of the star defined by the conserved mass as
\beqa
  a_{\ast} &\equiv& \Bigl( {M_{\ast} \over 4 \pi \r/3} \Bigr) ^{1/3} 
\nonumber \\
  &=&a_1 (\a_2 \a_3)^{1/3} \Bigl[ 1+ {1 \over c^2}
    \Bigl\{ {1 \over 30} F_a^2 \O_R^2
  (a_1^2+a_2^2) +{4 \over 5} \pi \r A_0 \Bigr\} \Bigr], 
\eeqa
and $a_{i \ast}$ defined by $a_{1 \ast}a_{2 \ast}a_{3 \ast}=a_{\ast}^3$, 
$a_{2 \ast}/a_{1 \ast}=a_2/a_1$ and $a_{3 \ast}/a_{1 \ast}=a_3/a_1$. 
$A_{ijk\cdots \ast}$ and $B_{ijk\cdots \ast}$ are 
computed by using  $a_{2\ast}/a_{1\ast}$ and $a_{3\ast}/a_{1\ast}$. 
In Eq. (\ref{ENPN}) we have used some relations among index
    symbols.\cite{chandra69}

\subsection{Total angular momentum}

The total angular momentum is written as
\beqa
  J &=& \int d^3 x \r \biggl[ v_{\varphi} \Bigl\{ 1+ {1 \over c^2}
  \Bigl( v^2 +6U +{P \over \r} \Bigr) \Bigr\} +{\hat{\b}_{\varphi}
    \over c^2} +O(c^{-4}) \biggr] \nonumber \\
  &=&J_{\rm N} +{1 \over c^2} J_{\rm PN}, \label{totangmom}
\eeqa
where
\beqa
  v_{\varphi} &=& -x_2 v_1 +x_1 v_2, \\
  \hat{\b}_{\varphi} &=& -x_2 \hat{\b}_1 +x_1 \hat{\b}_2, \\
  J_{\rm N}&=& {M_{\ast} \over 5} F_a \O_R (a_{1 \ast}^2 -a_{2 \ast}^2), \\
  J_{\rm PN}&=& J_{\rm N \ra PN} +J_{v_{\varphi} v^2} +J_{v_{\varphi} U}
  +J_{v_{\varphi} P/\r} +J_{\b_{\varphi}}, \\
  J_{\rm N \ra PN} &=&-J_{\rm N} \Bigl[ {1 \over 6} F_a^2 \O_R^2 
  (a_{1 \ast}^2 +a_{2 \ast}^2) +4 \pi \r A_{0 \ast} \Bigr] \nonumber \\
  & & +{M_{\ast} \over 5} (a_{1
    \ast}^2 -a_{2 \ast}^2) \Bigr( p +{3 \over 7} q a_{1 \ast}^2 +{3
    \over 7} r a_{2 \ast}^2 +{1 \over 7} s a_{3 \ast}^2 \Bigr) \nonumber \\
  & &+{M_{\ast} \over 5} F_a \biggl[ 2\O_R \Bigl\{ a_{1
    \ast}^2 S_{11} -a_{2 \ast}^2 S_{12} +{1 \over 7} a_{1 \ast}^2 S_{31}
  (a_{1 \ast}^2 +a_{2 \ast}^2) -{1 \over 7}
  a_{2\ast}^4 S_{32} \nonumber \\
  & & \hspace{40pt} -{1 \over 7} a_{1 \ast}^2 a_{3 \ast}^2 S_{33}
  \Bigr\} +\d \O (a_{1 \ast}^2
  -a_{2 \ast}^2) \biggr], \\
  J_{v_{\varphi} v^2} &=&{3M_{\ast} \over 35} F_a^3 \O_R^3 (a_{1 \ast}^4
  -a_{ 2\ast}^4), \\
  J_{v_{\varphi} U} &=&{12M_{\ast} \over 35} \pi \r F_a \O_R \Bigl[
   3A_{0 \ast} (a_{1 \ast}^2
  -a_{2 \ast}^2) -a_{1 \ast}^4 A_{1 \ast} +a_{2 \ast}^4 A_{2 \ast} \Bigr], \\
  J_{v_{\varphi} P/\r} &=&{2M_{\ast} \over 35} {P_0 \over \r} F_a \O_R
  (a_{1 \ast}^2 -a_{2 \ast}^2), \\
  J_{\b_{\varphi}} &=&{2M_{\ast} \over 35} \pi \r F_a \O_R (-a_{1 \ast}^2
  a_{2 \ast}^2 A_{1 \ast}
  -7a_{1 \ast}^4 A_{1 \ast} +a_{1 \ast}^2 a_{2 \ast}^2 A_{2 \ast} \nonumber \\
  & & \hspace{30pt} +7a_{2  \ast}^4 
  A_{2 \ast} -2a_{1 \ast}^4 a_{2 \ast}^2 A_{12 \ast} +2a_{1 \ast}^2
  a_{2 \ast}^4 A_{12 \ast}),
\eeqa
and
\beqa
  \d \O ={\d \O^2 \over 2\O_R}.
\eeqa


\section{Gravitational radiation}

In this section, we derive an equation for 
the luminosity of the gravitational radiation from a rotating star. 

\subsection{Frames}

In previous sections, we used the coordinate system $\{x_i\}$ 
associated with the principal axis of the ellipsoid, which is considered 
as the rotating frame. 
In these coordinates, the estimation of multipole moments is 
quite simple. 
On the other hand, it is convenient to use the inertial frame $\{X_i\}$ 
for the estimation of the luminosity of gravitational waves. 
These coordinate systems are related to each other as 
\begin{eqnarray}
X_1&=&x_1 \cos\Omega t-x_2 \sin\Omega t , \nonumber\\
X_2&=&x_1 \sin\Omega t+x_2 \cos\Omega t , \nonumber\\
X_3&=&x_3 .
\end{eqnarray}

In the inertial frame, the Newtonian internal velocity
is
\beq
v_i=(-\Omega F_1 X_2 ,~\Omega F_2 X_1 ,~0), \label{nvelo2}
\eeq
where we define $F_i$ for $i=1,2$ as 
\begin{equation}
F_i \equiv 1+{a_i^2 \over a_1^2+a_2^2} f_R. 
\end{equation}
Note that this expression of the Newtonian velocity is different from
Eq. (\ref{nvelo}) because we do not restrict our attention to the 
irrotational case in this section. 
If one takes $-2$ for $f_R$, Eq. (\ref{nvelo2}) reduces 
to Eq. (\ref{nvelo}).

\subsection{Multipole moments}

In the 1PN approximation, 
the energy loss due to gravitational radiation is 
written as\cite{bd89}
\begin{equation}
{dE \over dt}=-{1 \over 5c^5} \Bigl[ \sum_{i,j}M_{ij}^{(3)}M_{ij}^{(3)}
+{1 \over c^2} \Bigl({5 \over 189}\sum_{i,j,k} M_{ijk}^{(4)}M_{ijk}^{(4)}
+{16 \over 9}\sum_{i,j}S_{ij}^{(3)}S_{ij}^{(3)} \Bigr)\Bigr] , 
\label{edot}
\end{equation}
where $(n)$ denotes that $n$ time derivatives are taken, and 
$M_{ij}$, $M_{ijk}$ and $S_{ij}$ are written as 
\begin{eqnarray}
M_{ij}&\equiv&K_{ij}+V_{ij}+U_{ij}+P_{ij}+Y_{ij}+R_{ij}, \\
K_{ij}&\equiv&\int d^3X \rho \hat X_{ij} , \\
V_{ij}&\equiv&2\int d^3X \rho v^2 \hat X_{ij} , \\ 
U_{ij}&\equiv&2\int d^3X \rho U \hat X_{ij} , \\ 
P_{ij}&\equiv&3\int d^3X P \hat X_{ij} , \\ 
Y_{ij}&\equiv&{1 \over 14}{d^2 \over dt^2} \int d^3X \rho X^2 \hat X_{ij} , 
\\ 
R_{ij}&\equiv&-{20 \over 21}{d \over dt} \int d^3X \rho \sum_k 
v_k \hat X_{ijk} , \\
M_{ijk}&\equiv&\int d^3X \rho \hat X_{ijk} , \\
S_{ij}&\equiv&\int d^3X \rho \sum_{k,l} \epsilon_{kl<i}\hat X_{j>k}v_l , 
\end{eqnarray}
where $\epsilon_{kli}$ denotes the spatial Levi-Civita symbol, and 
both `hat' and $<\;>$ denote the symmetric traceless part; that is,
\begin{eqnarray}
  \hat X_{ij} &=& X_i X_j -{1 \over 3} \d_{ij} X^2, \\
  \hat X_{ijk} &=& X_i X_j X_k -{1 \over 5} X^2 (\d_{ij} X_k +\d_{jk} X_i
  +\d_{ki} X_j ), \\
  A_{<ij>}&=&{1 \over 2}(A_{ij}+A_{ji})-{1 \over 3}\delta_{ij}\sum_k A_{kk}. 
\end{eqnarray}

First, we consider the mass multipole moments associated with 
the principal axis of the ellipsoid. 
We obtain mass quadrupole moments to 1PN order as 
\begin{eqnarray}
I_{11}&\equiv&\int_V d^3x \rho x_1^2 \nonumber\\
&=&{4\pi \over 15} \rho a_1^3a_2a_3\Bigl[ 1+{2 \over c^2}
\Bigl( S_{11}+{a_1^2 \over 7} S_{31} -{a_3^2 \over 7} S_{33}
\Bigr)\Bigr]+O(c^{-4}) ,
\label{Ixx} \\
I_{22}&\equiv&\int_V d^3x \rho x_2^2 \nonumber\\
&=&{4\pi \over 15} \rho a_1a_2^3a_3\Bigl[ 1+{2 \over c^2}
\Bigl( S_{12}-{a_1^2 \over 7} S_{31} +{a_2^2 \over 7} S_{32}
\Bigr)\Bigr]+O(c^{-4}) ,
\label{Iyy} \\
I_{12}&\equiv&\int_V d^3x \rho x_1x_2=0. 
\label{Ixy} 
\end{eqnarray}
These moments are related to the reduced moments in the inertial frame as 
\begin{eqnarray}
K_{11}&=&{1 \over 2}(I_{11}-I_{22}) \cos 2\Omega t +{\rm const}, \\  
K_{22}&=&-K_{11}+{\rm const}, \\
K_{12}&=&{1 \over 2}(I_{11}-I_{22}) \sin 2\Omega t +{\rm const}. 
\end{eqnarray}

\subsection{Energy loss}

The quadrupole moments relevant to the energy loss are obtained as 
\begin{eqnarray}
V_{11}-V_{22}&=&\Omega^2 \Bigl[ -{1 \over 2}(I_{1111}+I_{2222}-6I_{1122}) 
(F_1^2-F_2^2) \cos 4\Omega t \nonumber\\
&&\hspace{20pt}+(I_{1111}-I_{2222}) (F_1^2+F_2^2) \cos 2\Omega t \Bigr]
+{\rm const} , \\
V_{12}&=&\Omega^2 \Bigl[ -{1 \over 4}(I_{1111}+I_{2222}-6I_{1122})
(F_1^2-F_2^2) \sin 4\Omega t \nonumber\\ 
&&\hspace{20pt}+{1 \over 2} (I_{1111}-I_{2222}) (F_1^2+F_2^2) \sin 2\Omega t 
\Bigr] +{\rm const} , \\
U_{11}-U_{22}&=&2\pi \rho \Bigl[ A_0 (I_{11}-I_{22})
-\sum_l A_l (I_{11ll}-I_{22ll}) \Bigr] \cos 2\Omega t , \\
U_{12}&=&\pi \rho \Bigl[ A_0 (I_{11}-I_{22})
-\sum_l A_l (I_{11ll}-I_{22ll}) \Bigr] \sin 2\Omega t , \\
P_{11}-P_{22}&=&{3P_0 \over \rho} \Bigl[ (I_{11}-I_{22})
-\sum_l {1 \over a_l^2} (I_{11ll}-I_{22ll}) \Bigr] 
\cos 2\Omega t , \\
P_{12}&=&{3P_0 \over 2\rho} \Bigl[ (I_{11}-I_{22})
-\sum_l {1 \over a_l^2} (I_{11ll}-I_{22ll}) \Bigr] 
\sin 2\Omega t , \\
Y_{11}-Y_{22}&=&-{2 \over 7}\Omega^2 (I_{1111}-I_{2222}+I_{1133}-I_{2233}) 
\cos 2\Omega t , \\
Y_{12}&=&-{1 \over 7}\Omega^2 (I_{1111}-I_{2222}+I_{1133}-I_{2233}) 
\sin 2\Omega t , \\
R_{11}-R_{22}&=&\Omega^2 \Bigl[ {20 \over 21} (F_1-F_2) 
(I_{1111}+I_{2222}-6I_{1122}) \cos 4\Omega t \nonumber\\
&&\hspace{10pt}-{8 \over 21}(F_1+F_2)(I_{1111}-I_{2222}+I_{1133}-I_{2233}) 
\cos 2\Omega t \Bigr], \\
R_{12}&=&\Omega^2 \Bigl[ {10 \over 21} (F_1-F_2) 
(I_{1111}+I_{2222}-6I_{1122}) \sin 4\Omega t \nonumber\\
&&\hspace{10pt}-{4 \over 21}(F_1+F_2)(I_{1111}-I_{2222}+I_{1133}-I_{2233}) 
\sin 2\Omega t \Bigr].
\end{eqnarray}

The higher multipole moments relevant to gravitational radiation 
in the 1PN approximation vanish due to the symmetry, i.e., 
\begin{eqnarray}
M_{ijk}=&&O(c^{-2}) , \\
S_{ij}=&&O(c^{-2}) . 
\end{eqnarray}

In total, we obtain 
\begin{eqnarray}
{dE \over dt}=-{32 \over 5c^5} (I_{11}-I_{22}) \Omega^6 
\biggl[ && (I_{11}-I_{22}) 
+{1 \over c^2} \Bigl\{ 
2 (I_{1111}-I_{2222}) (F_1^2+F_2^2) \Omega^2 \nonumber\\
&&+4\pi \rho \Bigl( A_0 (I_{11}-I_{22})-\sum_l A_l 
(I_{11ll}-I_{22ll}) \Bigr) \nonumber\\ 
&&+{6 P_0 \over \rho} \Bigl( I_{11}-I_{22}-\sum_l {1 \over a_l^2} 
(I_{11ll}-I_{22ll}) \Bigr) \nonumber\\
&&-\Bigl( {4 \over 7}+{16 \over 21}(F_1+F_2) \Bigr) 
(I_{1111}-I_{2222}+I_{1133}-I_{2233}) \Omega^2 \Bigr\}\biggr] \nonumber \\
&&+O(c^{-8}), 
\end{eqnarray}
where temporal averaging has been done. 

Using some algebraic relations among $I_{ii}$ and $I_{iiii}$, the formula 
for the energy loss can be expressed in terms of $(\rho, ~\Omega, 
~a_1, ~a_2, ~a_3, ~S_{11}, ~\cdots, ~S_{33}, ~A_0, ~f_R)$ as  
\begin{eqnarray} 
{dE \over dt}=&&-{32 \over 5c^5} \Bigl( {4\pi \over 15} \Bigr)^2 
( \rho a_1a_2a_3 )^2 (a_1^2-a_2^2) \Omega^6 \nonumber\\
&&\times\biggl[ (a_1^2-a_2^2) 
+{2 \over c^2} \Bigl\{ 2a_1^2 \Bigl( S_{11} +{a_1^2 \over 7} S_{31}
-{a_3^2 \over 7} S_{33} \Bigr)  \nonumber \\
&&\hspace{90pt}
-2a_2^2 \Bigl( S_{12} -{a_1^2 \over 7} S_{31} +{a_2^2 \over 7} S_{32}
\Bigr) \nonumber\\
&&\hspace{10pt}+(a_1^2-a_2^2) \Bigl( {38 \over 21} \pi \rho A_0 
+{1 \over 147} {\Omega_R^2 \over a_1^2+a_2^2} 
{\cal F}(f_R,a_1,a_2,a_3) \Bigr) \Bigr\}\biggr] , 
\label{edot2}
\end{eqnarray}
where we define ${\cal F}(f_R,a_1,a_2,a_3)$ as 
\begin{eqnarray}
{\cal F}(f_R,a_1,a_2,a_3)\equiv&&
(63f_R^2+102f_R+11)(a_1^4+a_2^4) \nonumber\\
&&-(49f_R^2-92f_R-22)a_1^2a_2^2 -(8f_R+22)a_3^2(a_1^2+a_2^2) . ~~~~~
\end{eqnarray}

\subsection{The Energy loss rate for irrotational Riemann
  ellipsoids}

For a 1PN irrotational ellipsoid $(f_R=-2)$,\cite{chandra69}
we obtain 
\begin{eqnarray} 
{dE \over dt}=&&-{32 \over 5c^5} \Bigl( {4\pi \over 15} \Bigr)^2 
( \rho a_1a_2a_3 )^2 (a_1^2-a_2^2) \Omega^6 \nonumber\\
\times\biggl[ && (a_1^2-a_2^2) 
+{2 \over c^2} \Bigl[ 2a_1^2 \Bigl( S_{11} +{a_1^2 \over 7} S_{31}
-{a_3^2 \over 7} S_{33} \Bigr) 
-2a_2^2 \Bigl( S_{12} -{a_1^2 \over 7} S_{31} +{a_2^2 \over 7} S_{32}
\Bigr) \nonumber\\
& &\hspace{40pt} +(a_1^2-a_2^2) \Bigl\{ {38 \over 21} \pi \rho A_0 \nonumber \\
& & \hspace{40pt}+{1 \over 147} 
\Bigl( 59(a_1^4+a_2^4)-358 a_1^2a_2^2 -6a_3^2 (a_1^2+a_2^2) \Bigr) 
{\Omega_R^2 \over a_1^2+a_2^2} \Bigr\} \Bigr] \biggr] .  
\label{edotirrot}
\end{eqnarray}


We can rewrite Eq. (\ref{edotirrot}) by using the conserved mass as
\beqa
  {dE \over dt} = \Bigl( {dE \over dt} \Bigr)_{\rm N} +{1 \over c^2}
  \Bigl( {dE \over dt} \Bigr)_{\rm PN}, \label{dedtirre}
\eeqa
where
\beqa
  \Bigl( {dE \over dt} \Bigr)_{\rm N} &=&-{32 M_{\ast}^2 \over 125 c^5}
  (a_{1 \ast}^2 -a_{2 \ast}^2)^2 \O_R^6, \\
  \Bigl( {dE \over dt} \Bigr)_{\rm PN} &=&-{64 M_{\ast}^2 \over 125c^5}
  (a_{1 \ast}^2 -a_{2 \ast}^2) \O_R^6 \nonumber \\
&\times &\biggl[ 2a_{1 \ast}^2 \Bigl( S_{11}
  +{a_{1 \ast}^2 \over 7} S_{31} -{a_{3 \ast}^2 \over 7} S_{33} \Bigr)
  -2a_{2 \ast}^2 \Bigl( S_{12} -{a_{1 \ast}^2 \over 7} S_{31} +{a_{2
      \ast}^2 \over 7} S_{32} \Bigr) \nonumber \\
  & &\hspace{1pt}+(a_{1 \ast}^2-a_{2 \ast}^2)
  \Bigl\{ {3 \d \O^2 \over 2\O_R^2} -{46 \over 21} \pi \r A_{0 \ast} -{1 
    \over 6} F_a^2 \O_R^2 (a_{1 \ast}^2 +a_{2 \ast}^2) \nonumber \\
  & &\hspace{1pt}+{1 \over 147}
  \Bigl( 59(a_{1 \ast}^4+a_{2 \ast}^4)-358 a_{1 \ast}^2 a_{2 \ast}^2
  -6a_{3 \ast}^2 (a_{1 \ast}^2+a_{2 \ast}^2) \Bigr) {\Omega_R^2 \over
    a_{1 \ast}^2 +a_{2 \ast}^2} \Bigr\} \biggr]. \nonumber \\
\eeqa
Note that in Eq. (\ref{dedtirre}) there are no terms dependent on the 
1PN velocity, since the quadrupole moment $M_{ij}$ in Eq. (\ref{edot}) 
does not depend on the velocity.

\section{Numerical results}

\subsection{Normalization}

First of all, we change to non-dimensional parameters according to
\beqa
  &&\t{p} \equiv {p \over \pi \r \L a_1^2},~~~\t{q} \equiv {q \over \pi \r 
    \L},~~~\t{r} \equiv {r \over \pi \r \L},~~~\t{s} \equiv {s \over \pi
    \r \L}, \nonumber \\
  &&\t{S}_{11} \equiv {S_{11} \over \pi \r a_1^2},~~~\t{S}_{12} \equiv
  {S_{12} \over \pi \r a_1^2},~~~\t{S}_{31} \equiv {S_{31} \over \pi \r}
,~~~\t{S}_{32} \equiv {S_{32} \over \pi \r},~~~\t{S}_{33} \equiv
  {S_{33} \over \pi \r},
\eeqa
and also
\beqa
  \t{\O} \equiv {\O \over \sqrt{\pi \r}},~~~\d \t{\O}^2 \equiv {\d \O^2
    \over (\pi \r a_1)^2},~~~\d \t{\O} \equiv {\d \O \over (\pi \r)^{3/2}
      a_1^2}.
\eeqa
Using these parameters with the relation $S_{11} =S_{12}$, we can rewrite
equations (\ref{cond1}), (\ref{cond10}),
(\ref{cond2}) $\sim$ (\ref{cond4}), and (\ref{cond5}) $\sim$ (\ref{cond7})
into the non-dimension forms. From these eight equations, we can
determine eight variables, $\t{p}$, $\t{q}$, $\t{r}$, $\t{s}$,
$\t{S}_{11}$, $\t{S}_{31}$, $\t{S}_{32}$ and $\t{S}_{33}$.

After determination of the variables, we can calculate the energy and
angular momentum of the star from Eqs. (\ref{totenergy}) and
(\ref{totangmom}) and also the luminosity of gravitational radiation
from Eq. (\ref{edotirrot}). In the numerical calculations these
values must be normalized. The normalized energy, angular
momentum and luminosity of gravitational radiation are written as
\beqa
  \t{E} &\equiv& {E \over M_{\ast}^2/a_{\ast}}
  =\t{E}_{\rm N} +C_{\rm s} \t{E}_{\rm PN}, \label{tE} \\
  \t{J} &\equiv& {J \over (M_{\ast}^3 a_{\ast})^{1/2}}
  =\t{J}_{\rm N} +C_{\rm s} \t{J}_{\rm PN}, \label{tJ} \\
  \t{d E \over dt} &\equiv& \Bigl( {dE \over dt} \Bigr) \Big/ 
\Bigl({M_{\ast}^5
    \over c^5 a_{\ast}^5}\Bigr)
  =\Bigl( \t{dE \over dt} \Bigr)_{\rm N} +C_{\rm s} \Bigl(
  \t{dE \over dt} \Bigr)_{\rm PN}, \label{tdEdt}
\eeqa
where $C_{\rm s}$ is the compactness parameter defined as
\beqa
  C_{\rm s} \equiv {M_{\ast} \over c^2 a_{\ast}}.
\eeqa
In the PN approximation, we assume $ C_{\rm s} \ll 1$. 

\subsection{Ellipsoidal approximation}

The ellipsoidal approximation, in which the equilibrium configuration is
assumed to have an ellipsoidal shape, is useful in order to investigate
features of stars or binary systems in the 1PN
approximation,\cite{TS,Lom,SZ} because the whole calculation is done by
setting $S_{ij}=0$, and hence is very simplified. The ellipsoidal
approximation gives an exact solution for the rotating incompressible
star in the Newtonian theory. However, it gives only an approximate
solution for the 1PN case. If the ellipsoidal approximation is really an
excellent approximation, it becomes a robust method for the study of the
1PN effects. This is motivation for our investigation of the validity of
the ellipsoidal approximation.

\begin{table}
 \caption{The Newtonian and 1PN 
angular velocity, energy, angular momentum, and 
luminosity of gravitational radiation along the
equilibrium sequence of the 
irrotational Riemann S-type ellipsoid.
}
 \begin{center}
  {\tabcolsep=2pt
  \begin{tabular}{cl|lll|cll|cc} \hline \hline
    $a_2/a_1$&$a_3/a_1$&$~~\t{\O}_R^2$&$~~\t{E}_{\rm N}$&$~~\t{J}_{\rm N}$&
    $\d \t{\O}^2$&$~~\t{E}_{\rm PN}$&$~~\t{J}_{\rm PN}$&
    $(d \tilde{E}/dt)_{\rm N}$&$(d \tilde{E}/dt)_{\rm PN}$ \\ \hline
    0.99&0.9950&0.2667&-0.6000&1.807(-5)&
    8.050(-2)&-4.284(-2)&2.957(-5)&-8.275(-7)&5.153(-6) \\
    0.95&0.9741&0.2667&-0.5998&4.707(-4)&
    7.837(-2)&-4.235(-2)&7.312(-4)&-2.160(-5)&1.341(-4) \\
    0.90&0.9464&0.2669&-0.5991&1.986(-3)&
    7.776(-2)&-4.071(-2)&2.879(-3)&-9.184(-5)&5.660(-4) \\
    0.85&0.9167&0.2673&-0.5979&4.728(-3)&
    7.936(-2)&-3.775(-2)&6.359(-3)&-2.214(-4)&1.348(-3) \\
    0.80&0.8848&0.2677&-0.5960&8.917(-3)&
    8.301(-2)&-3.324(-2)&1.106(-2)&-4.258(-4)&2.546(-3) \\
    0.75&0.8505&0.2683&-0.5934&1.483(-2)&
    8.841(-2)&-2.694(-2)&1.682(-2)&-7.269(-4)&4.244(-3) \\
    0.70&0.8136&0.2688&-0.5898&2.282(-2)&
    9.515(-2)&-1.861(-2)&2.338(-2)&-1.156(-3)&6.544(-3) \\
    0.65&0.7736&0.2692&-0.5851&3.332(-2)&
    0.1026~~~~&-8.211(-3)&3.006(-2)&-1.757(-3)&9.594(-3) \\
    0.60&0.7304&0.2691&-0.5789&4.691(-2)&
    0.1102~~~~&~3.220(-3)&3.425(-2)&-2.591(-3)&1.370(-2) \\
    0.55&0.6835&0.2683&-0.5711&6.433(-2)&
    0.1188~~~~&~1.989(-3)&1.035(-2)&-3.739(-3)&2.110(-2) \\
    $\dagger$ 0.5244&0.6579&0.2673&-0.5663&7.505(-2)&
            $\infty$~~~&~~$\infty$&~~~$\infty$&-4.481(-3)&$\infty$ \\
    0.50&0.6325&0.2660&-0.5612&8.660(-2)&
    0.1111~~~~&~9.301(-2)&0.1499&-5.305(-3)&1.530(-2) \\
    0.45&0.5772&0.2614&-0.5485&0.1151&
    0.1117~~~~&~0.1021&0.1357&-7.403(-3)&2.337(-2) \\
    0.40&0.5174&0.2534&-0.5324&0.1516&
    0.1025~~~~&~0.1379&0.1699&-1.012(-2)&2.763(-2) \\
    0.35&0.4531&0.2403&-0.5121&0.1988&
    8.362(-2)&~0.1895&0.2423&-1.345(-2)&2.898(-2) \\
    0.30&0.3850&0.2206&-0.4864&0.2606&
    5.094(-2)&~0.2762&0.4161&-1.707(-2)&2.507(-2) \\
    0.25&0.3143&0.1927&-0.4540&0.3428&
    -7.291(-2)~~&~0.7824&1.632&-2.018(-2)&-3.537(-3)~ \\
    $\ddagger$ 0.2374&0.2963&0.1842&-0.4447&0.3678&
             $\infty$~~~&~~~$\infty$&~~~$\infty$&~2.071(-2)&$\infty$ \\
    0.20&0.2430&0.1559&-0.4135&0.4551&
    6.538(-2)&~5.760(-3)&1.493(-3)&-2.127(-2)&2.552(-2) \\
    0.15&0.1737&0.1117&-0.3630&0.6154&
    2.253(-2)&~0.1183&0.8808&-1.861(-2)&1.067(-2) \\
    0.10&0.1094&6.454(-2)&-0.2994&0.8666&
    5.964(-3)&~0.1163&2.973&-1.172(-2)&2.893(-3) \\
    0.05&5.167(-2)&2.252(-2)&-0.2137&1.370&
    5.969(-4)&~6.554(-2)&14.44&-3.460(-3)&9.916(-5) \\ \hline
  \end{tabular}}
 \end{center}
\end{table}%

In the ellipsoidal approximation, a solution is obtained by setting
$S_{ij}=0$ in all the equations. After we set $S_{ij}=0$, we can
calculate the velocity field in the 1PN approximation from Eqs.
(\ref{cond1}) and (\ref{cond2}) $\sim$ (\ref{cond4}). We note that in the
ellipsoidal approximation, the boundary condition (\ref{pressurecond})
is not satisfied.

\subsection{Results}

We give the results in Tables I, II and III. In Table I, the angular
velocity, energy, angular momentum, and luminosity of gravitational
radiation of Newtonian and 1PN orders are presented along the Newtonian
sequence. Coefficients of the 1PN velocity potential and the Lagrangian
displacement vectors are shown in Table II. In these tables, $\dagger$
and $\ddagger$ denote singularities at which all the 1PN terms diverge.
We believe that at those points, some instabilities which are concerned
with the fourth order harmonics of the Riemann ellipsoid are induced by
relativistic corrections.

\begin{table}
 \caption{Coefficients of 
the 1PN velocity potential and 
the Lagrangian displacement vectors 
shown along the equilibrium sequence of
the irrotational Riemann S-type ellipsoid.
}
 \begin{center}
  {\tabcolsep=3pt
  \begin{tabular}{c|clll|llll} \hline \hline
    $a_2/a_1$&$\t{p}$&~~~~$\t{q}$&~~~~$\t{r}$&~~~~$\t{s}$&
    ~~~~$\t{S}_{11}$&~~~~$\t{S}_{31}$&~~~~$\t{S}_{32}$&
    ~~~~$\t{S}_{33}$ \\ \hline
    0.99&~~~3.337(-2)&-1.893(-3)&-1.936(-3)&-1.914(-3)&
         -9.723(-6)&~4.082(-5)&~8.268(-6)&-2.560(-6) \\
    0.95&0.1635&-9.213(-3)&-1.034(-2)&-9.786(-3)&
         -2.393(-4)&~1.017(-3)&~2.096(-4)&-5.569(-5) \\
    0.90&0.3180&-1.781(-2)&-2.245(-2)&-2.021(-2)&
         -9.335(-4)&~3.996(-3)&~8.393(-4)&-1.655(-4) \\
    0.85&0.4623&-2.593(-2)&-3.643(-2)&-3.147(-2)&
         -2.033(-3)&~8.639(-3)&~1.841(-3)&-1.926(-4) \\
    0.80&0.5947&-3.403(-2)&-5.209(-2)&-4.389(-2)&
         -3.470(-3)&~1.425(-2)&~3.050(-3)&~1.025(-4) \\
    0.75&0.7137&-4.334(-2)&-6.855(-2)&-5.794(-2)&
         -5.159(-3)&~1.929(-2)&~4.050(-3)&~1.133(-3) \\
    0.70&0.8177&-5.704(-2)&-8.293(-2)&-7.450(-2)&
         -7.021(-3)&~2.019(-2)&~3.860(-3)&~3.615(-3) \\
    0.65&0.9051&-8.443(-2)&-8.605(-2)&-9.551(-2)&
         -9.071(-3)&~7.270(-3)&-4.744(-5)&~8.805(-3) \\
    0.60&0.9768&-0.1620&-4.032(-2)&-0.1273&
         -1.192(-2)&-5.589(-2)&-1.615(-2)&~1.885(-2) \\
    0.55&1.060~~~&-0.6220&~0.4026&-0.2216&
         -2.342(-2)&-0.4951&-0.1118&~3.530(-2) \\
    $\dagger$ 0.5244&$\infty$&~~~~$\infty$&~~~~$\infty$&~~~~$\infty$&
            ~~~~$\infty$&~~~~$\infty$&~~~~$\infty$&~~~~$\infty$ \\
    0.50&0.9322&~0.9441&-1.289&-1.317(-2)&
         ~1.656(-2)&~1.043&~0.1764&~9.268(-2) \\
    0.45&0.9589&~0.4818&-0.8558&-0.1228&
         ~8.354(-3)&~0.5938&~4.776(-2)&~0.1726 \\
    0.40&0.9099&~0.4682&-0.9012&-0.1789&
         ~1.439(-2)&~0.5780&-4.266(-2)&~0.3572 \\
    0.35&0.8081&~0.5694&-1.073&-0.2463&
         ~2.877(-2)&~0.6705&-0.2372&~0.8163 \\
    0.30&0.6232&~0.8617&-1.440&-0.3617&
         ~6.274(-2)&~0.9574&-0.8477&~2.320 \\
    0.25&-0.2187~~&~3.306&-3.834&-0.9421&
         ~0.3245&~3.571&-7.001&~18.19 \\
    $\ddagger$ 0.2374&$\infty$&~~~~$\infty$&~~~~$\infty$&~~~~$\infty$&
             ~~~~$\infty$&~~~~$\infty$&~~~~$\infty$&~~~~$\infty$ \\
    0.20&0.8031&-0.6934&-0.2521&-0.2627&
         -9.644(-2)&-0.8229&~3.568&-9.625 \\
    0.15&0.4650&-8.768(-2)&-1.024&-0.6262&
         -2.999(-2)&-0.2098&~2.168&-6.409 \\
    0.10&0.2583&~7.056(-2)&-1.555&-1.133&
         -1.064(-2)&-6.378(-2)&~1.877&-6.251 \\
    0.05&0.1006&~0.1021&-2.758&-2.463&
         -2.574(-3)&-1.361(-2)&~1.885&-7.055  \\ \hline
  \end{tabular}}
 \end{center}
\end{table}%

As shown in Table II, the velocity field of 1PN order has an 
$x_3$ component which is written as
\beq
  v_3 =\p_3 \phi_{\rm PN}= 2 s x_1 x_2 x_3,
\eeq
where we use Eq. (\ref{phipn}). This feature has also been found in the
study of the Dedekind ellipsoid.\cite{chandra74} In the 1PN (and
relativistic) case, we believe that the velocity field cannot be
restricted in the planes orthogonal to the figure rotation axis in
general.

In Table III, the angular velocity, energy, angular momentum, luminosity
of gravitational radiation and velocity field of 1PN order in the
ellipsoidal approximation are presented. There appear no singularities
which are associated with instabilities because we do not take into account
deformation of the ellipsoidal figure. From Table III, it is found that
the velocity field of 1PN order has an $x_3$ component, as in the case of
the exact calculation. This implies that it is necessary to include the
$x_3$ component of the velocity field even in the ellipsoidal
approximation.

\begin{table}
 \caption{The 1PN angular velocity, energy, angular momentum,
luminosity of gravitational radiation and velocity field 
in the ellipsoidal approximation along the
equilibrium sequence of the irrotational Riemann S-type ellipsoid. The
subscript ${\rm e}$ denotes the case of the {\it ellipsoidal}
approximation.
}
 \begin{center}
  {\tabcolsep=4pt
  \begin{tabular}{c|lll|c|llll} \hline \hline
    $a_2/a_1$&~~~$\d \t{\O}^2_{\rm e}$&~~~$\t{E}_{\rm PN e}$&
    ~~~$\t{J}_{\rm PN e}$&
    $(d \tilde{E}/dt)_{\rm PN e}$&
    ~~~$\t{p}_{\rm e}$&~~~$\t{q}_{\rm e}$&~~~$\t{r}_{\rm e}$&
    ~~~$\t{s}_{\rm e}$ \\ \hline
    0.99&8.050(-2)&-4.284(-2)&2.954(-5)&5.154(-6)&
         3.336(-2)&-1.921(-3)&-1.907(-3)&-1.914(-3) \\
    0.95&7.826(-2)&-4.235(-2)&7.275(-4)&1.343(-4)&
         0.1633&-9.959(-3)&-9.595(-3)&-9.787(-3) \\
    0.90&7.735(-2)&-4.073(-2)&2.849(-3)&5.676(-4)&
         0.3172&-2.088(-2)&-1.938(-2)&-2.021(-2) \\
    0.85&7.844(-2)&-3.780(-2)&6.259(-3)&1.354(-3)&
         0.4608&-3.287(-2)&-2.948(-2)&-3.149(-2) \\
    0.80&8.138(-2)&-3.336(-2)&1.084(-2)&2.562(-3)&
         0.5929&-4.601(-2)&-4.012(-2)&-4.388(-2) \\
    0.75&8.592(-2)&-2.713(-2)&1.645(-2)&4.276(-3)&
         0.7120&-6.030(-2)&-5.165(-2)&-5.774(-2) \\
    0.70&9.168(-2)&-1.884(-2)&2.296(-2)&6.596(-3)&
         0.8169&-7.563(-2)&-6.466(-2)&-7.353(-2) \\
    0.65&9.812(-2)&-8.131(-3)&3.027(-2)&9.633(-3)&
         0.9056&-9.168(-2)&-8.001(-2)&-9.187(-2) \\
    0.60&0.1045&~5.362(-3)&3.835(-2)&1.350(-2)&
         0.9762&-0.1079&-9.899(-2)&-0.1136 \\
    0.55&0.1099&~2.202(-2)&4.729(-2)&1.826(-2)&
         1.026&-0.1232&-0.1234&-0.1398 \\
    0.50&0.1130&~4.218(-2)&5.749(-2)&2.388(-2)&
         1.054&-0.1360&-0.1560&-0.1720 \\
    0.45&0.1126&~6.598(-2)&6.993(-2)&3.009(-2)&
         1.055&-0.1439&-0.2003&-0.2122 \\
    0.40&0.1071&~9.315(-2)&8.678(-2)&3.621(-2)&
         1.026&-0.1438&-0.2611&-0.2633 \\
    0.35&9.567(-2)&~0.1226&0.1130&4.096(-2)&
         0.9658&-0.1320&-0.3441&-0.3289 \\
    0.30&7.828(-2)&~0.1519&0.1599&4.251(-2)&
         0.8710&-0.1049&-0.4562&-0.4147 \\
    0.25&5.664(-2)&~0.1763&0.2559&3.900(-2)&
         0.7430&-6.093(-2)&-0.6049&-0.5289 \\
    0.20&3.434(-2)&~0.1886&0.4789&2.994(-2)&
         0.5874&-3.273(-3)&-0.8009&-0.6868 \\
    0.15&1.596(-2)&~0.1799&1.078&1.763(-2)&
         0.4161&~5.746(-2)&-1.070&-0.9242 \\
    0.10&4.719(-3)&~0.1427&3.088&6.679(-3)&
         0.2469&~0.1024&-1.512&-1.357 \\
    0.05&4.973(-4)&~7.597(-2)&14.52&9.189(-4)&
         9.926(-2)&~0.1057&-2.718&-2.594 \\ \hline
  \end{tabular}}
 \end{center}
\end{table}%

In Fig.~1, we show the angular velocity of Newtonian and 1PN orders as a
function of the axial ratio $a_2/a_1$. Figures 1(a) and (b) are drawn
for the Newtonian solutions, and (c) and (d) are for the 1PN solutions.
The solid and dotted lines denote the results of the exact calculation
and those of the ellipsoidal approximation, respectively. In the
Newtonian case (or in the spherical case), the above two lines coincide.
On the other hand, in the 1PN (and non-spherical) cases, the differences
between the two lines appear, but are not large.

The energy and angular momentum of Newtonian and 1PN orders are shown in
Fig.~2 as functions of the axial ratio $a_2/a_1$. In Fig.~2, we present
the results only for $1 \ge a_2/a_1 \ge 0.55$, i.e., for the stable
branch.  The solid and dotted lines denote the same quantities as those
in Fig.~1.  The luminosity of gravitational radiation is shown in
Fig.~3.  The solid and dotted lines are again the same as in Fig.~1.
From the figures, it is found that with increasing the 1PN correction,
the angular velocity and angular momentum increase, while the total
energy and luminosity of gravitational waves decrease. These features
are reasonable, because with increasing relativistic correction, the
self-gravity of the star becomes strong (and hence the binding energy
becomes large), and to support the self-gravity, angular velocity must
be larger.

\begin{figure}[t]
\epsfysize 14cm 
\begin{center}
\leavevmode
\epsfbox{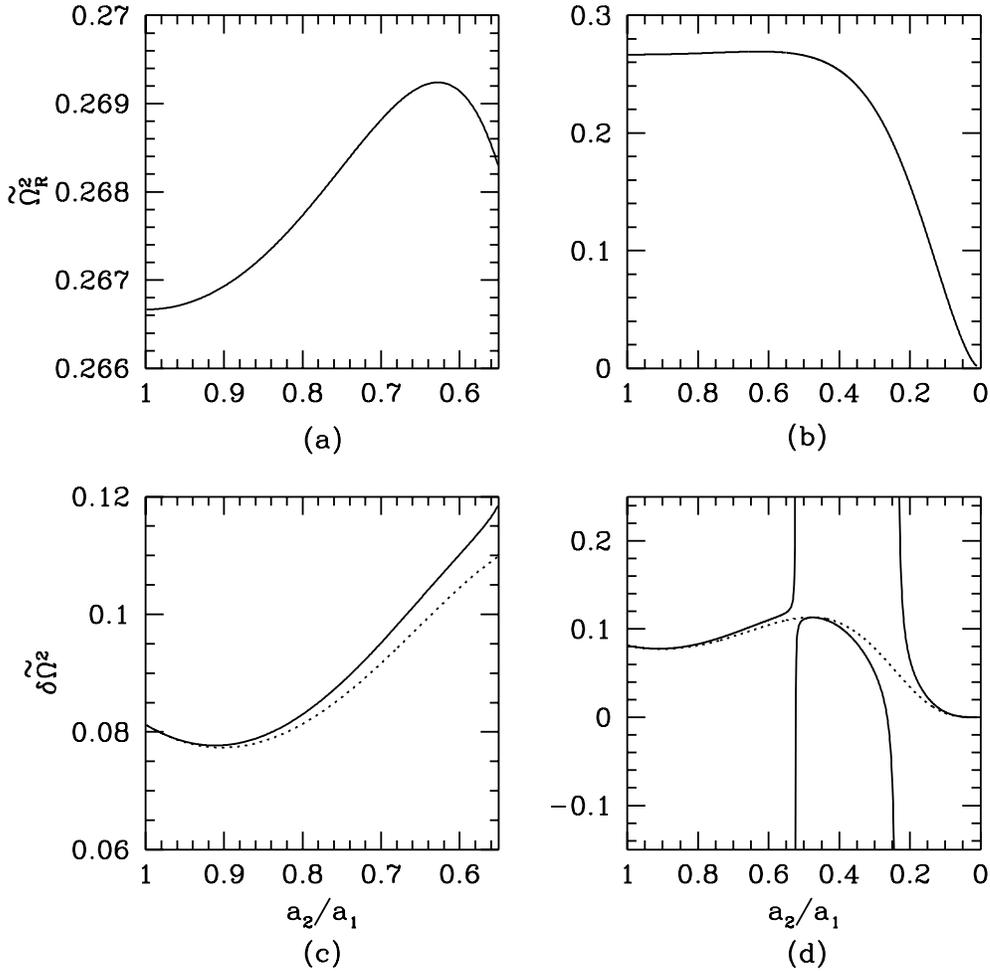}
\end{center}
\caption{The angular velocity of Newtonian (a) and (b) and 1PN (c) and (d)
orders as functions of the axial ratio $a_2/a_1$. Figures 1(a)
and (c) are magnifications of Figs. 1(b) and
(d). The solid and dotted lines denote the results of the exact 
calculation and of the ellipsoidal approximation, respectively. 
}
\end{figure}

\section{Summary and discussion}

We have calculated an equilibrium sequence and the luminosity of
gravitational waves of an irrotational and incompressible star in the
1PN approximation. We have developed a scheme to solve the hydrostatic
equations for an irrotational fluid in the 1PN approximation. By
extending the present study, we will investigate irrotational binary
stars in the 1PN approximation. Also, the 1PN solutions presented in
this paper will be useful when examining the accuracy of numerical code
for obtaining relativistic irrotational stars.

\begin{figure}[t]
\epsfysize 14cm 
\begin{center}
\leavevmode
\epsfbox{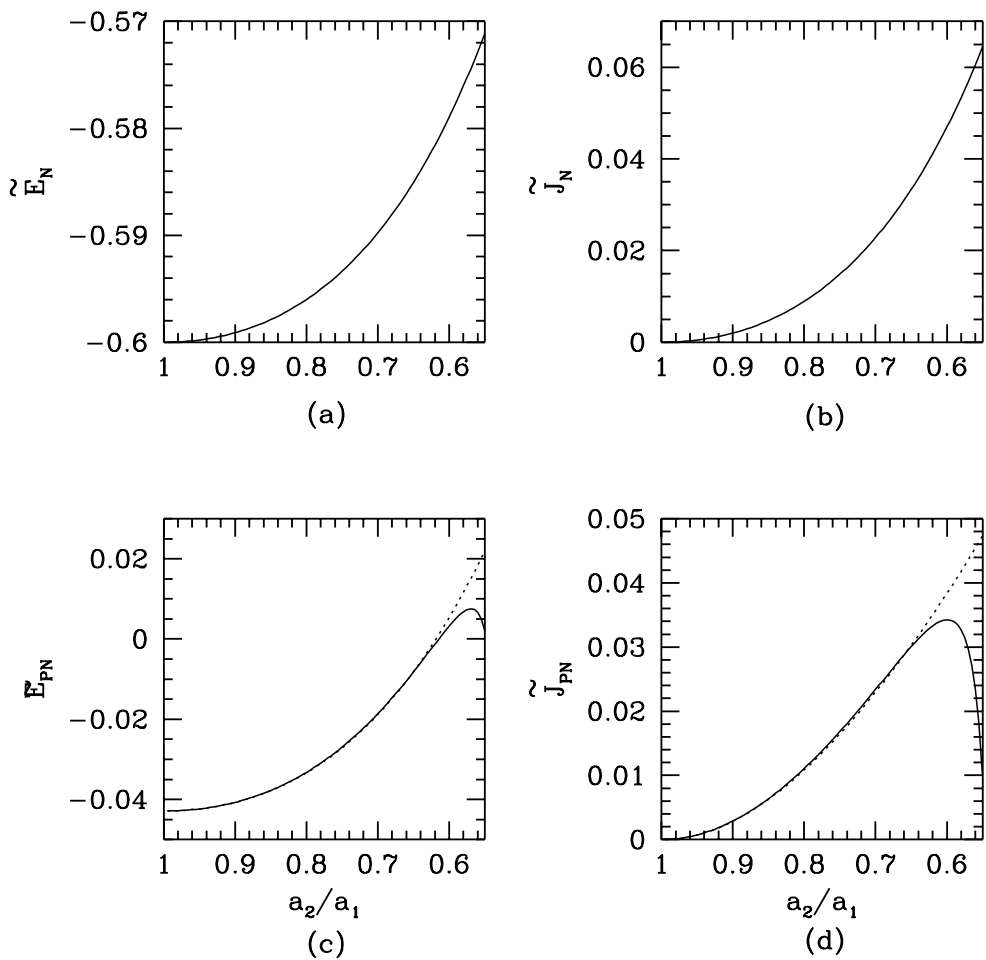}
\end{center}
\caption{The total energy of Newtonian (a) and 1PN (c) orders, and the
total angular momentum of Newtonian (b) and 1PN (d) orders
as a function of the axial ratio $a_2/a_1$. The solid and dotted lines
are the same as in Fig. 1.
}
\end{figure}

In addition to establishing the method, we have an interesting result.
The 1PN velocity field has an $x_3$ component. This fact implies that
internal motion never remains in the planes orthogonal to the figure
rotation axis. (Note that such a waving motion closes even if it has an
$x_3$ component.) We can find in the Chandrasekhar and Elbert paper
\cite{chandra74} that there is also an $x_3$ component of the internal
motion in the case of the 1PN Dedekind ellipsoid. Moreover, Ury\=u and
Eriguchi\cite{UE} numerically showed that the internal flow of the
compressible Dedekind-like star has an $x_3$ component even in the
Newtonian case. Therefore, except for {\it incompressible}, {\it
  Newtonian} cases, non-axisymmetric stars with an internal motion will
have an $x_3$ component of the velocity field in general.

\begin{figure}[t]
\epsfysize 14cm
\begin{center}
\leavevmode
\epsfbox{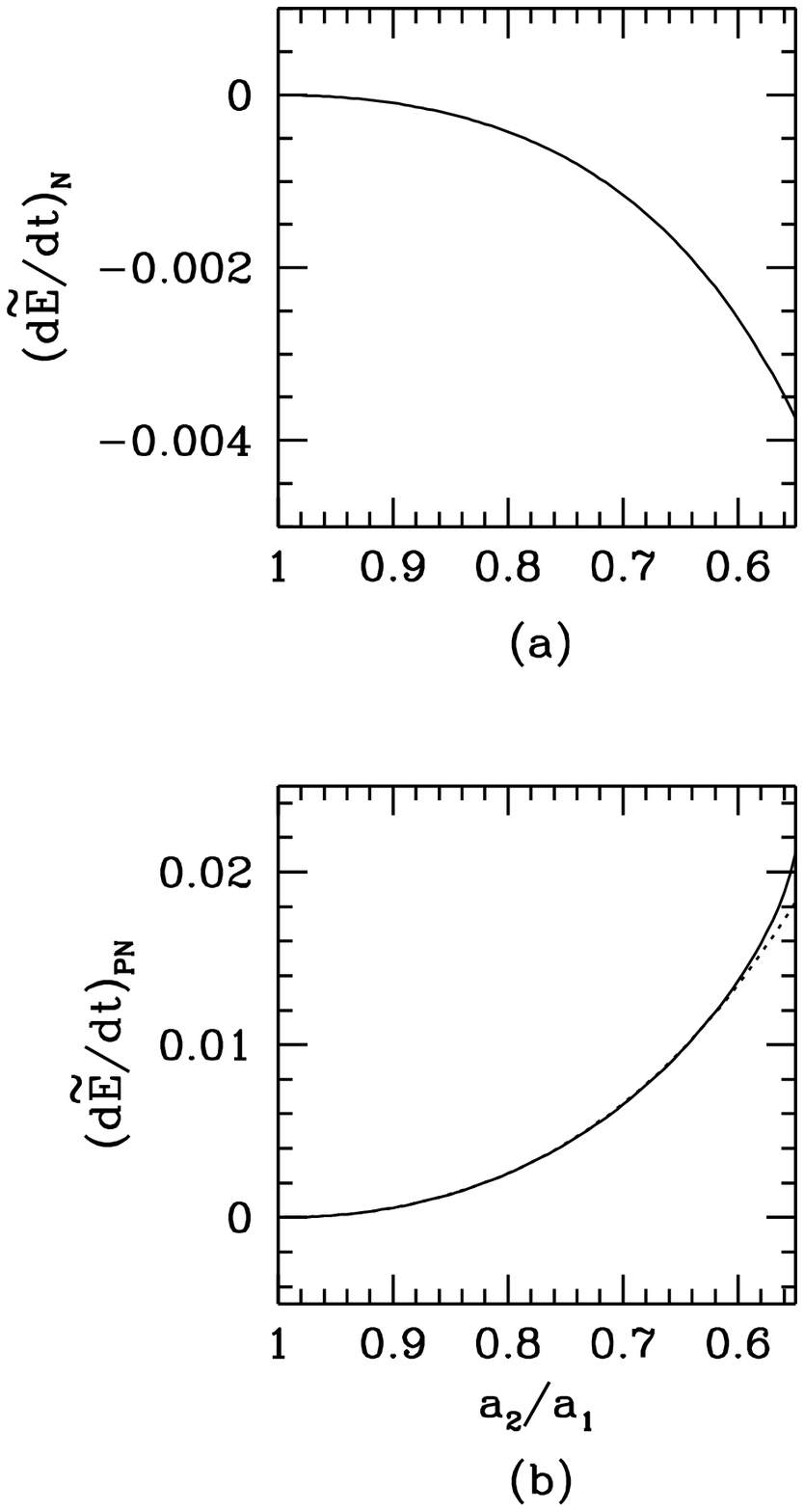}
\end{center}
\caption{The luminosity of gravitational radiation of the Newtonian (a)
  and 1PN (b) orders as functions of the axial ratio $a_2/a_1$. The
  solid and dotted lines are the same as in Fig. 1.
}
\end{figure}

Finally, we discuss the validity of the ellipsoidal approximation.  One
can see from Figs.~1 $\sim$ 3 and Tables I $\sim$ III that the error of the
1PN angular velocity, energy, and angular momentum in the ellipsoidal
approximation are less than about $\sim 5\%$ in the range of $1 \ge
a_2/a_1 \ge 0.7$. Although some of the coefficients of the 1PN velocity
potential, $\t{q}$ and $\t{r}$, deviate from exact values by a large
factor near $a_2/a_1 \simeq 0.55$, the error is not so large for larger
$a_2/a_1$.  Therefore, we may conclude that the ellipsoidal
approximation is a good one if the star is not highly nonspherical ($1
\ge a_2/a_1 \ge 0.7$). This result is encouraging for study of binary
stars in the ellipsoidal approximation because in the binary case, it is
expected that $a_2/a_1$ and $a_3/a_1$ do not become very small.
\cite{TS}

\section*{Acknowledgements}

We thank Y. Eriguchi for helpful comments. 
K. T. would like to thank T. Nakamura and H. Sato for helpful comments
and continuous encouragement. This work was partly supported by a
Grant-in-Aid for Scientific Research Fellowship (No.9402) and a
Grant-in-Aid (Nos. 08NP0801 and 09740336) of the Japanese Ministry of
Education, Science, Sports and Culture.

\end{document}